\documentclass[aps,pra,epsf,twocolumn]{revtex4}
\usepackage{graphicx}
\usepackage{subfigure}
\usepackage{bm}

\usepackage[intlimits]{amsmath} 

\usepackage{verbatim} 
\usepackage{psfrag}

\usepackage{amssymb}
  \newcommand{\ket}[1]{ | #1 \rangle  }

\newcommand{\V}[1]{{\bf #1}}

\newcommand{\expct}[1]{\langle #1 \rangle}

\newcommand{\rmi}{{\rm i}}
\newcommand{\rme}{{\rm e}}
\newcommand{\rmd}{{\rm d}}

\begin{document}

\title{Homodyne detection of matter-wave fields}

\author{Stefan Rist$^{1,2,3}$ and Giovanna Morigi$^{1,3}$}

\affiliation{
$^1$ Departament de F{\'i}sica, Universitat Aut\`onoma de Barcelona, 08193
Bellaterra, Spain\\
$^2$ NEST, Istituto Nanoscienze-CNR and Scuola Normale Superiore, I-56126 Pisa, Italy\\
$^3$ Theoretische Physik, Universit\"at des Saarlandes, D-66041 Saarbr\"ucken,
Germany}

\begin{abstract}
A scheme is proposed, that allows one for performing homodyne detection of the
matter-wave field of ultracold bosonic atoms. It is based on a pump-probe lasers setup, that both illuminates a Bose-Einstein condensate, acting as reference system, and a second ultracold gas, composed by the same atoms but in a quantum phase to determine. Photon scattering outcouples atoms from both systems, which then propagate freely. Under appropriate conditions, when the same photon can either be scattered by the Bose-Einstein condensate or by the other quantum gas, both flux of outcoupled atoms and scattered photons exhibit oscillations, whose amplitude is proportional to the condensate fraction of the quantum gas. This setup allows one, for instance, to perform thermometry of a condensate or to monitor the Mott-insulator/superfluid phase transition in optical lattices, and can be  extended in order to measure the first-order correlation function of a quantum gas. The dynamics here discussed make use of the entanglement between atoms and photons, which is established by the scattering process, in order to access detailed information on the quantum state of matter. \end{abstract}

\date{\today}
\maketitle

\section{Introduction} 

\begin{figure}[htp] \centering
\includegraphics[width=0.34\textwidth]{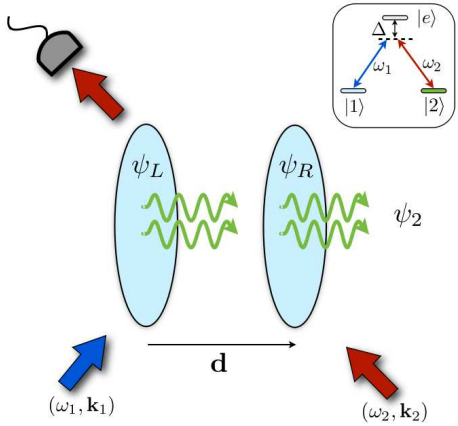}
\caption{ \label{fig:HomodyneSetup} Setup for performing homodyne detection of
matter-wave fields. A gas of identical, bosonic atoms is confined in two distinguishable regions of space. In the left region it forms a Bose-Einstein condensate, that acts as reference system for measuring the condensate fraction of the right system. A pump-probe laser setup outcouples the atoms from both wells. The photon flux of the probe beam is constituted by the photon scattered in the outcoupling process and exhibits oscillations whose amplitude depend on the condensate fraction of the right system, and which vanish when this is zero. The first-order correlation function of a quantum gas can be determined in an extension of this setup, where the shaded regions denoted by $\psi_L$ and $\psi_R$ correspond to the spatial regions of the quantum gas trapped in a single well. In this case, the amplitude of the photon flux oscillations provides a measurement of the spatial correlation function as a function of the distance $d$. Detail on the parameters are reported in the text.
}
\end{figure}

The measurement technique, which is mostly employed in experiments with
ultracold atoms, is based on time of flight \cite{Stringari_RMP,Bloch-Review}.
Used in different setups, it allows one to determine the quantum state 
of matter \cite{Demler_Lectures,Bach,Zwerger}. The most recent
demonstration of in-situ imaging of atoms in optical lattices
\cite{Greiner_Science_2010,Bloch_Nature_2010} has opened further exciting
directions for probing and manipulating atoms in periodic arrays
\cite{Greiner_PRL2011,Bloch_Science_2011}. In addition, the remarkable progress in coupling ultracold atomic gases with high-finesse optical resonators \cite{Reichel,Esslinger,Stamper-Kurn,Courteille} has generated a renewed interest in revealing the correlation functions of matter by photodetection of the scattered light. This progress opens, amongst several others, the possibility to monitor properties of the quantum state of matter in a non-destructive way \cite{Meystre,RitschNature07,LarsonPRL,Mekhov_QND}. 

Photons emitted by Rayleigh scattering, and in general in a pump-probe type of experiment,  have been used in order to characterize the quantum state of atomic gases \cite{Hemmerich,Phillips,Zimmermann_Bragg,LightScattMottBloch,Davidson_RMP}. They deliver information on the structure form factor of the quantum gas and thus on the density and density-density correlations \cite{Davidson_RMP,Demler_Lectures,Rist2010,Douglas11}.  Proposals for optical detection of certain quantum states of matter have appeared in the literature \cite{Lewenstein93,Javanainen,Ruotekoski09,Rist2010,Douglas11}. To our knowledge, however, a setup has not yet been identified, which would allow one measure the mean value of the atomic field operator by means of the scattered light. This measure would  be analogous to homodyne detection of light fields \cite{Homodyne},  and specifically allow one to determine the condensate fraction of a quantum gas \cite{Stringari_Book} by photodetection.

In this article we propose a setup in which detection of the photons scattered by a quantum gas allows one to determine the mean value of the atomic field.  The setup for determining the condensate fraction is sketched in Fig. \ref{fig:HomodyneSetup}, and is based on an interferometer for Bose-Einstein Condensates (BEC) realized in Refs.~\cite{Saba_DetPhase,Shin_OptWeakLinkBEC}. We establish a direct connection between our theoretical model and the experimental setup in Ref.~\cite{Saba_DetPhase,Shin_OptWeakLinkBEC}, and then focus on a scheme in which the second system is not necessarily a BEC. We then show that this setup allows one to determine the condensate fraction of the second system by measuring the flux of the scattered photon. Furthermore, we argue that this scheme can be extended to determine the first-order correlation function of an atomic gas, when the two illuminated regions belong to the same quantum gas. The dynamics is based on first creating entanglement between the scattered photon and the scattering atom of the quantum gas, and then on performing an operation similar to a quantum eraser \cite{scully_QuantumEraser}, thereby allowing one to measure the correlation functions of matter in the flux of the scattered light. 

This article is organized as follows. In Sec. \ref{Sec:1} the setup and the
theoretical model are introduced. In Sec. \ref{Sec:2} a formal connection is derived between 
the photon flux and the correlation functions of the scattering atoms. 
In Sec. \ref{Sec:3} the photon flux is evaluated for three
exemplary cases: when the scattering systems are two Bose-Einstein condensates
at zero temperature, when they are at different temperatures, and when one of
the systems is composed by ultracold atoms in a two-dimensional optical lattice.
In Sec. \ref{Sec:4} it is discussed how this setup may allow one to measure
the first-order correlation function. The conclusions are drawn in Sec. \ref{Sec:5}. The appendices report details of the calculations presented in Secs. \ref{Sec:2}, \ref{Sec:3}, and \ref{Sec:4}.

\section{The Model}
\label{Sec:1}

A realization of the setup we consider in this paper is depicted in Fig.~\ref{fig:HomodyneSetup}. Here, ultracold bosonic atoms of mass $m$ are initially prepared in the stable electronic state $\ket {1}$ and confined by the (state-dependent) potential $V_1({\bf r})$. The atoms, that are illuminated by the lasers, are coherently coupled to the electronic state $\ket{2}$, which is stable and not confined: the outcoupled atoms propagate freely in space. 

In this section we introduce the Hamiltonian and the physical quantities which
are at the basis of our analysis. 

\subsection{Hamiltonian}

We denote by $H$ the Hamiltonian governing the dynamics of photons and atoms,
which we decompose into the sum 
\begin{equation}
H=H_0+H_{\rm emf}+H_{\rm int}+H_{\rm shift}\,, 
\end{equation}
with $H_0$ the Hamiltonian governing the atoms dynamics in absence of
interaction with the lasers, $H_{\rm emf}$ the Hamiltonian for the free,
transverse electromagnetic (e.m.) field, while $H_{\rm int}$ describes the
coupling between states $\ket{1}$ and $\ket{2}$ induced by coherent Raman scattering. 
Hamiltonian $H_{\rm shift}$
includes the dynamical Stark shift due to off-resonant coupling between atoms
and e.m.-field modes. These Hamiltonians are reported in a second-quantized
description of matter- and photon-fields. In detail, $$H_{\rm
emf}=\sum_{\lambda}\hbar\omega_\lambda a_\lambda^{\dagger}a_\lambda\,$$
gives the energy of the transverse electromagnetic field in free space (and is
reported without the vacuum energy), where $\lambda$ labels a mode of the e.m.-field at 
wave vector ${\bf k}$, polarization $\vec{\epsilon}\perp {\bf k}$,
and frequency $\omega_{\bf k}=c|{\bf k}|$, with $c$ the velocity of light. The
operators $a_{\lambda}$ and $a_{\lambda}^{\dagger}$ annihilate and create,
respectively, a photon in mode $\lambda$, and obey the bosonic commutation
relation $[a_\lambda,a_{\lambda'}^\dagger]=\delta_{\lambda,\lambda'}$. 

The atomic Hamiltonian $H_0$ is conveniently rewritten as $H_0=H_1+H_2+H_{12}$,
where
\begin{eqnarray} 
H_{\{j=1,2\}} &=& \int {\rm d}\V{r}\,\psi_j ^{\dagger }(\V{r}%
)\left( -\frac{\hbar ^{2}\nabla ^{2}}{2m}+V_j(\V{r}) +\hbar\omega_{aj}\right)
\psi_j(\V{r}) \nonumber \\ && +\frac{g_j}{2}\int {\rm d}\V{r}\psi_j
^{\dagger }(\V{r})\psi_j ^{\dagger }(\V{r}%
)\psi_j (\V{r})\psi_j (\V{r})\, \label{eqn:HamOutcouplGen:1}
\end{eqnarray}
describes the atoms dynamics when the atoms are in the electronic state $\ket{j}$
(at frequency $\omega_{aj}$). Here $\psi_j({\bf r})$ and $\psi_j^{\dagger}({\bf
r})$ are the annihilation and creation operator, respectively for a bosonic atom
at position ${\bf r}$ and in the electronic state $\ket{j=1,2}$, obeying the
commutation relation
$[\psi_j(\V{r}),\psi_k^\dagger(\V{r}')]=\delta_{j,k}\delta(\V{r}-\V{r}')$. Parameter
$g_j=4 \pi \hbar^2 a_{s,j}/m$ is the strength of $s$-wave scattering between the atoms in state
$\ket{j}$, with $a_{s,j}$ the $s$-wave
scattering length in the corresponding electronic state \cite{Stringari_Book}.
Hamiltonian term $H_{12}$ describes $s$-wave scattering between an atom in
electronic state $\ket{1}$ and an atom in 
electronic state $\ket{2}$: We do not report its explicit form as we will consider situations for which these types of collisions can be neglected. The interested reader can find the corresponding term, for instance, in
Refs. \cite{Maciej,QTweezers}.

The Hamiltonian terms describing the interactions between
photons and atoms are given using normal ordering \cite{Larson_NJP}. They include a term describing coherent Raman 
coupling between the states $\ket{1}$ and  $\ket{2}$, in which a photon in mode $1$ is scattered into mode $2$ and vice versa. Raman transitions follow from a pump-probe type of excitation, and the corresponding Hamiltonian reads
\begin{equation}\label{eqnKet:V:0}
H_{\rm int} = \hbar \int {\rm d}\V{r}\left[\gamma(\V{r})a_{2}^{\dagger
}\psi _{2}^{\dagger }(\V{r})\psi _{1}(
\V{r})a_{1}+{\rm H.c.}\right] \, ,
\end{equation}
with the  position-dependent coupling strength $\gamma(\V{r})$. This interaction can be tailored by means of a laser, pumping the atoms, and a cavity
mode, acting as a probe, that are both far-detuned from the atomic excited state
but are set close to resonance with an atomic Raman transition. For later convenience we denote by the
quantity $$\omega_{12}=\omega_{1}-\omega_2$$ the difference between the
frequencies of the two e.m.f.-modes. Moreover, we assume
$\omega_{12}\sim\omega_{a2}-\omega_{a1}$, i.e., the two lasers drive
quasi-resonantly the Raman transition coupling the stable electronic states
$\ket{1}$ and  $\ket{2}$.
The interaction with the pump and probe laser also induces a dynamical Stark shift, whose corresponding Hamiltonian reads (in normally-ordered form)
\begin{equation}
H_{\rm shift}=\sum_{j=1}^2\hbar\int {\rm d}{\bf r} \gamma_j({\bf
r})a_{j}^{\dagger}\psi _{j}^{\dagger }(\V{r})\psi _{j}(\V{r})a_{j}\,,
\end{equation}
where the parameter $\gamma_j({\bf r})$ has the dimension of an angular
frequency. It can be written as $\gamma_j({\bf r})=2|\Omega_j({\bf r})|^2/\Delta$, with
$\Omega_j({\bf r})$ the Rabi frequency of the mode coupling the transition
$\ket{j}\to \ket{e}$ and $\Delta$ the detuning between the mode and the atomic
transition frequencies (the Lamb shifts are included in the frequencies of the atomic states).

In the following we will assume that the pump and probe fields are traveling waves
with wave vectors ${\bf k_j}$. In this case, the Raman-coupling strength
$\gamma({\bf r})$ reads
\begin{equation}
\gamma({\bf r})=\gamma_0{\rm e}^{{\rm i}{\bf q}\cdot {\bf r}}
\end{equation}
with $\V{q}=\V{k}_1-\V{k}_2$ and $\gamma_0=2\Omega_1({\bf r})\Omega_2({\bf
r})^*/\Delta$.

\subsection{Spin-dependent potential}

We will now provide more details on the potential which confines the atoms. 
We will assume that the atoms in state $\ket{2}$ propagate
freely, namely, $V_2(\V{r})=const$ in Eq. (\ref{eqn:HamOutcouplGen:1}), which we
set equal to zero. 

The atoms in state $\ket{1}$  are trapped by potential $V_1(\V{r})$. This
potential confines the atoms in two spatially separated regions around the two
potential minima, which are at distance $d$ and, specifically, are localized at
the points ${\bf r_L}=-(d/2) \hat x$ and ${\bf r_R}=(d/2) \hat x$. The potential
can be then decomposed in the sum $V_1({\bf r})=V_L(\V{r})+V_R(\V{r})$, with
$V_j(\V{r})$ the potential center at ${\bf r_j}$ and $j=L,R$. The two atomic
clouds at each well are initially uncorrelated, and there is no tunneling
between the two spatial region. 

In principle, hence, the atoms in the two regions of space are distinguishable. 
It is then useful to consider the partition $\psi_1({\bf r})=\psi_L({\bf
r})+\psi_R({\bf r})$, where we denote by $\psi_L({\bf r},t)=\psi_1({\bf
r},t)\theta(-z)$ and $\psi_R({\bf r},t)=\psi_1({\bf r},t)\theta(z)$ the field
operators which do not vanish on the left and right region of space,
respectively, such that
$[\psi_j(\V{r}),\psi_k^\dagger(\V{r}')]=\delta_{j,k}\delta(\V{r}-\V{r}')$ for
$\V{r},\V{r}'\neq 0$ \cite{comrel}. Using these definitions, we can write $H_1=H_L+H_R$, with
\begin{eqnarray} 
H_{\{j=L,R\}} &=& \int {\rm d}\V{r}\,\psi_j ^{\dagger }(\V{r}%
)\left( \frac{-\hbar ^{2}\nabla ^{2}}{2m}+V_j(\V{r}) \right)
\psi_j(\V{r}) \nonumber \\ && +\frac{g}{2}\int {\rm d}\V{r}\psi_j
^{\dagger }(\V{r})\psi_j ^{\dagger }(\V{r}%
)\psi_j (\V{r})\psi_j (\V{r})\, ,\label{eqn:HamOutcouplGen:a}
\end{eqnarray}
and where we have set $\omega_{a1}=0$. With this representation, the Hamiltonian describing the
interaction with the lasers takes the form
\begin{equation}\label{eqnKet:V}
H_{\rm int} = \hbar \int {\rm d}\V{r}\left[\gamma(\V{r})a_{2}^{\dagger}\psi
_{2}^{\dagger }(\V{r})\left(\psi _{L}(
\V{r})+\psi _{R}(\V{r})\right)
a_{1}+{\rm H.c.}\right] \,.
\end{equation}
We note that
our model is an extension to the one studied in Refs. \cite{LuxatOutcoupling,Choi_Outcoupling}, where the authors studied the outcoupling of atoms from a single BEC by means of {\it classical} Raman
lasers. By considering the quantum dynamics of the interaction between photons and atoms, we take into account the quantum fluctuations of the light field due to the scattering process, which are discarded in \cite{LuxatOutcoupling,Choi_Outcoupling}. We will indeed show that these fluctuations give access to some correlation functions of the scattering system.

\subsection{The scattered field}

We now study the properties of the scattered photons by considering the photon field operators in
Heisenberg picture. The Heisenberg equations of motion are determined assuming that the coupling between matter and photons is sufficiently weak to be treated in second-order perturbation theory. In this limit the operators for the photonic modes read
\begin{widetext}
\begin{eqnarray}
\label{eqn:Heisenberg:1}
a_1(t)&=&a_1(0) {\rm e}^{-{\rm i}(\omega_1 t+\phi_1(t))}-{\rm i}\,{\rm e}^{-{\rm
i}\omega_1 t}\int_0^t {\rm d}\tau\, {\rm e}^{{\rm i}\omega_{12} \tau} \int {\rm
d}{\bf r}\;\gamma({\bf r}) \Bigl[\psi_L^{(0)\dagger}({\bf
r},\tau)+\psi_R^{(0)\dagger}({\bf r},\tau)\Bigr] \psi_2^{(0)}({\bf r},\tau)
a_2(0) \, ,\\
\label{eqn:Heisenberg:2}
a_2(t)&=&a_2(0) {\rm e}^{-{\rm i}(\omega_2 t+\phi_2(t))}-{\rm i}\,{\rm e}^{-{\rm
i}\omega_2 t}\int_0^t {\rm d}\tau\, {\rm e}^{-{\rm i}\omega_{12}\tau} \int {\rm
d}{\bf r}\;\gamma({\bf r})  \psi_2^{(0)\dagger}({\bf
r},\tau)\Bigl[\psi_L^{(0)}({\bf r},\tau)+\psi_R^{(0)}({\bf r},\tau)\Bigr] a_1(0)
\, ,
\end{eqnarray}
\end{widetext}
where $\psi_j^{(0)}({\bf r},\tau)=\exp({\rm i}H_0\tau/\hbar)\psi_j({\bf
r},0)\exp(-{\rm i}H_0\tau/\hbar)$. The physical origin of the individual terms on the
Right-Hand Side (RHS) can be simply identified. The first term on the RHS of both equations is the
free-field component. It is characterised by a time-dependent phase $\phi_j(t)$, which is proportional to the atomic density in the electronic state $\ket{j}$. When the Born approximation can be performed, it is a
density-dependent phase shift of the field mode, with the form

\begin{eqnarray*}
&&\phi_1(t)=\sum_{j,k=L,R}\int_0^t{\rm d}\tau\int{\rm d}{\bf r}\;\gamma_1({\bf
r})\psi _{j}^{(0)\dagger }(\V{r},\tau)\psi^{(0)}_{k}(\V{r},\tau)\,,\\
&&\phi_2(t)=\int_0^t{\rm d}\tau\int{\rm d}{\bf r}\;\gamma_2({\bf r})\psi
_{2}^{(0)\dagger }(\V{r},\tau)\psi _{2}^{(0)}(\V{r},\tau)\,.
\end{eqnarray*}
Since all atoms are initially prepared in the electronic state $\ket{1}$, this
density-dependent shift can be measured in the frequency shift the field-mode $1$ after it has interacted with the atomic medium. This would allow a quantum-non-demolition measurement of the density of the medium integrated over the path along which light propagates \cite{Demler_Lectures}. It is interesting to note that in the considered setup this shift contains also the operator $\psi _{L}^{(0)\dagger}(\V{r},\tau)\psi_{R}^{(0)}(\V{r},\tau)$ and its adjoint. Hence, its measurement would reveal tunneling events between the wells. This is however not relevant
for the case considered in this work, since we assume that there is no tunneling between the two separate wells. 

The second term on the RHS establishes a direct proportionality relation between the photonic  and the matter-wave fields $a_2$ and $\psi_1({\bf r})$ ($a_1$ and $\psi_2({\bf r})$). It shows, in particular, that the source term of field $a_1$ is the coherent overlap of the field scattered from the right and from the left wells. In this shape the outcoupling process in this system resembles a beam-splitter operation. This property establishes an analogy with an interferometric setup, which we will exploit in order to perform homodyne detection of matter-wave field. Differing from a simple interferometer, however, atoms and photons are correlated by the scattering process. This important difference is at the basis of the dynamics we observe. 

Before we discuss the signal at the photodetector, we will introduce few
approximations which will notably simplify the treatment. In first place we neglect atomic collisions
between outcoupled atoms in state $\ket{2}$, since we assume
that the gas of outcoupled atoms is at very low densities. This regime also
permits us to neglecting collisions between outcoupled atoms and trapped atoms
in state $\ket{1}$ \cite{intout}.

\section{Photon Flux}
\label{Sec:2}

The quantity we analyse in this paper is the flux of photons of mode 2, namely,
the rate of change of the photon number in the mode at frequency $\omega_2$. 
Formally, the photon flux is given by the equation
\begin{equation}\label{Ftdef}
 F(t)=\frac{d}{{\rm d}t}\expct{a_2^\dagger (t) a_2(t)} \, 
\end{equation}
where the expectation value $\expct{\cdot}$ is taken over the initial state of
the atoms and the e.m.-field. The photon flux, integrated over the
detection time, gives the integrated intensity of the field at the detector. It can be
verified that
\begin{equation} \label{eqn:MLequiv}
F(t) \propto\frac{\rmd}{\rmd t} \int {\rm d} \V{r} \expct{\psi_2^\dagger(\V{r},t)
\psi_2(\V{r},t)}
\end{equation}
where the proportionality factor is the mean number of photons in mode 1. This
equality shows indeed that flux of scattered photons and of the corresponding
outcoupled atoms carry the same information.

For the specific setup we consider the photon flux can be rewritten as the sum
of two contributions, 
\begin{equation}
\label{photon:flux}
F=F_B+F_I\,,
\end{equation}
with
\begin{subequations}
\begin{eqnarray}
F_B(t) &=& \Gamma {\rm Re} \sum _{j=R,L}\int {\rm d}{\bf r}\int {\rm d}{\bf r'}
\nonumber \\ && \times \int_0^{t}{\rm d}t' f({\bf r},t;{\bf r'},t') G_{jj}({\bf
r},t;{\bf r'},t')\,,\label{F:B}\\
F_I(t) &=& \Gamma {\rm Re}\int {\rm d}{\bf r}\int {\rm d}{\bf r'}\int_0^{t}{\rm
d}t' f({\bf r},t;{\bf r'},t') \label{F:I}\nonumber \\ 
&&\times (G_{LR}({\bf r},t;{\bf r'},t')+G_{RL}({\bf r},t;{\bf r'},t'))\,,
\end{eqnarray}
\end{subequations}
where $\Gamma=2\expct{a_1^\dagger a_1}\gamma_0^2$ is a scaling factor,
proportional to the number of photons in mode 1. Here,
\begin{eqnarray}
&&G_{jk}({\bf r},t;{\bf r'},t')=\left\langle \psi_j^{\dagger}({\bf
r'},t')\psi_k({\bf r},t)\right\rangle \, ,\label{corr:1} \\
&&f({\bf r},t;{\bf r'},t')={\rm e}^{{\rm i}[{\bf q}\cdot ({\bf r}-{\bf
r'})-\omega_{12}(t-t')]}\left\langle \psi_2({\bf r'},t') \psi_2^{\dagger}({\bf
r},t)\right\rangle \, , \label{fun:f} \nonumber \\
\end{eqnarray}
where the initial state is the vacuum state of the e.m. field except for mode 1 and 2,
which are assumed to be in coherent states with non-vanishing photon number, while all atoms are in internal
state $|1\rangle$ and at equilibrium in the grand-canonical ensemble at
temperature $T$. The atoms in state $\ket{2}$ propagate freely and do not undergo collisions, since we assume that the density is very low. Therefore, Eq.~(\ref{fun:f}) can be cast in the form
\begin{equation}
\label{fun:f_final}
f({\bf r},t;{\bf r'},t')=  \int \frac{{\rm d} \V{k}}{(2 \pi)^3} {\rm e}^{{\rm i}[({\bf
q}-{\bf k}) \cdot ({\bf r}-{\bf
r'})-(\tilde{\omega}_{12}-\omega_{\V{k}})(t-t')]} \, ,
\end{equation}
with $$\omega_{\bf k} =\frac{\hbar {\bf k}^2}{2 m}$$ the recoil frequency and
$$\tilde{\omega}_{12}=\omega_{12}-\omega_{a2}$$ the two-photon detuning. 

In order to obtain explicit expressions for the correlation functions
$G_{jk}(\V{r},t;\V{r}',t')$ it is convenient to work in the interaction picture
with respect to the grand canonical ensemble $K_0=H_0-\sum_{j=L,R} \mu_j {\cal
N}_j$, where $\mu_j$ is the chemical potential of the atoms in either the right
or left cloud and ${\cal N}_j=\int {\rm d}{\bf r}\;\psi_j^{\dagger}({\bf
r})\psi_j({\bf r})$. The new atomic field operators are obtained from the ones
in Heisenberg picture with respect to $H_0$ by replacing
$\psi_j(\V{r},t)\rightarrow \psi_j(\V{r},t) \rme^{-\frac{\rmi}{\hbar}\mu_j t}$. 
From now on we will denote by $\psi_j(\V{r},t)$ the atomic field
operators in Heisenberg picture with respect to $K_0$.

We now discuss how the considered setup can be used in order to measure the mean value of the atomic field operator by means of photons. We first note that the photon flux is the sum of two terms: the component $F_B$, that is the sum of the flux from each well, and the component $F_I$, that arises from the coherent superposition of a photon (atom) scattered by the wells. We will denote $F_B$ by ``background contribution'' and $F_I$ by ``interference contribution''. In absence of initial correlations, this latter term is proportional to the product of the mean value of the field operators $\langle\psi_L({\bf r},t)\rangle\langle\psi_R({\bf r},t)\rangle^*$. Let us now consider the case in which, say, the left well confines weakly-interacting atoms forming a
Bose-Einstein condensate. Under this assumption the atomic field operator can be written as \cite{Stringari_Book}
\begin{equation} \label{eqn:PsiSplit}
\psi _{L}(\V{r},t) =\rme^{-\frac{\rmi}{\hbar}\mu_L t}\left(\Phi _{L}(\V{r})
+\delta \psi_{L}(\V{r},t)\right)\, ,
\end{equation}
where $\Phi _{L}(\V{r})$ is the macroscopic wave function, which solves the Gross-Pitaevskii equation for
the quantum gas in the left well of the potential (with chemical potential $\mu_L$). Then, when the order parameter of the left condensate is known, the interference term will deliver the mean value of the field operator in the right well. 

In the following we use Eq.~(\ref{eqn:PsiSplit}) in the equations for the background contribution to the photon flux, Eq. (\ref{F:B}), and for the interference term, Eq. (\ref{F:I}). In particular, we will take
\begin{equation}
\Phi _{L}(\V{r})=\expct{\psi_j(\V{r},t)}=f_L(\V{r}) e^{{\rm i} \varphi_L}
\end{equation}
and assume that both $f_L(\V{r})$ and $\varphi_L$ are real valued. Hence, $f_L(\V{r})^2$ is the density of condensed atoms, while the phase $\varphi_L$ is assumed to be constant in space (hence discarding the possibility of superfluid currents in the Bose-Einstein condensate \cite{Stringari_Book}). 

\subsection{Background contribution}

The background contribution can be decomposed into the sum of the photon flux from the left and from the right well, $F_B=F_L+F_R$. Since we make a specific assumption on the quantum state of the gas trapped in the left well, we can provide an explicit form for $F_L$. For this purpose, we consider the integrand $G_{LL}(\V{r},t;\V{r'},t')$ in $F_L$ and observe that it can be splitted into two terms:
\begin{eqnarray}
G_{LL}(\V{r},t;\V{r'},t')={\rm e}^{{\rm
i}\mu_L(t-t')/\hbar}\left[f_{L}(\V{r})^2+\delta
G_{LL}(\V{r},t;\V{r'},t')\right]\,,\nonumber\\
\end{eqnarray}
where 
\begin{eqnarray}
&&\delta G_{LL}(\V{r},t;\V{r'},t')=\langle
\delta\psi_L^{\dagger}(\V{r},t)\delta\psi_L\V{r'},t')\rangle\nonumber\\
&&=\int \frac{ {\rm d}\omega }{2\pi }{\rm e}^{-{\rm i}\omega(t-t')}N_0(\omega)
A_{\delta \psi_L \delta \psi_L^\dagger}(\V{r},\V{r}',-\omega)  
\end{eqnarray}
accounts for the contribution of the noncondensed atoms to the photon flux.
Here,
\begin{equation}\label{eqn:SingleParticle}
A_{\delta\psi\delta\psi^{\dagger }}(\V{r},\V{r}',\omega )
=\int_{-\infty }^{\infty }{\rm d}\tau\,e^{{\rm i} \omega \tau}\left\langle
\left[ \delta\psi
(\V{r},\tau),\delta\psi ^{\dagger }(\V{r}',0)\right] \right\rangle  \, 
\end{equation}
is the spectral density \cite{menotti2008}, with $N_{0}(\omega )=[e^{\beta \hbar \omega }-1]^ {-1}$.
Similar expressions can be found for the background contribution of the quantum
gas in the right well. 

In order to gain insight on the equations we just derived, we assume that the
gas can be considered homogeneous. 
This assumption allows us to write the spectral density as $A_{\delta \psi_j\delta
\psi_j^{\dagger }}(\V{r},\V{r}',\omega )=A_{\delta \psi_j\delta \psi_j^{\dagger
}}(|\V{r}-\V{r}'|,\omega )$. We then inspect the equations for the flux using
the momentum representation for the atomic field operator. The terms giving the
contribution to the photon flux from the left cloud can be cast in the form \cite{corsright}
\begin{subequations}\label{eqnKet:FCB} 
\begin{eqnarray}
F_L(t) &=& 2\pi \Gamma {\rm Re} \int \frac{{\rm d}\V{k}}{(2\pi)^3} |\tilde f_L(\V{k})|^2
\tilde\delta^{t}(\Omega-\omega_{\V{k}+\V{q}}) \, ,\\
\delta F_{L}(t) &=& 2\pi \Gamma V {\rm Re}  \int \frac{{\rm d}\omega }{2\pi } \int 
\frac{{\rm d}\V{k}}{(2\pi)^3}   \\ && \times N_0(\omega)
\tilde A_{\delta \psi_L \delta \psi_L^\dagger}(\V{k},-\omega)
\tilde\delta^{t}(\omega
+\Omega-\omega_{\V{k}+\V{q}})\, , \nonumber
\end{eqnarray}
\end{subequations}
where $\tilde A_{\delta \psi_R \delta \psi_R^\dagger}(\V{k},\omega)$ is the
spectral weight function, which is the Fourier transform of $A_{\delta \psi_L
\delta \psi_L^\dagger}(\V{r},\omega)$ and gives strength of the collective
excitations at wavevector $\V{k}$ and frequency $\omega$, weighted by the Bose
function $N_0(\omega)$ \cite{menotti2008}. The other parameters are the volume $V$ in which the atoms of the left well are confined, the Fourier transform of the macroscopic wave function $\tilde f_L(\V{k})=\int {\rm d}\V{r} {\rm e}^{-{\rm i} \V{k}\cdot \V{r}} f_L(\V{r})$, and the frequency
\begin{eqnarray}
\Omega &=& \tilde{\omega}_{12}-\frac{\mu_L}{\hbar} \, ,
\end{eqnarray}
giving the effective detuning of the laser from the collective transition. Moreover, in Eqs. \eqref{eqnKet:FCB} we have introduced the quantity
\begin{eqnarray}
\label{tilde:delta}
\tilde\delta^{t}(\omega)={\rm e}^{-{\rm i}\omega t/2}\delta^t(\omega)\,,
\end{eqnarray}
that is proportional to the diffraction function $\delta^t(\omega)=\sin (\omega t/2) /(\pi\omega)$, enforcing energy conservation for long interaction times $t$. In particular, $\tilde\delta^{t}(\omega)\to\delta(\omega)$ for $t\to\infty$,  with $\delta(\omega)$ Dirac-delta function \cite{Cohen}. Definition \eqref{tilde:delta} contains a
time-dependent phase and
highlights that the corresponding factor tends to unity in the limit in which
energy conservation applies. Nevertheless, since we are going to also consider
intermediate times, we will systematically keep this time-dependent phase in the equations for the photon flux.

The integrals in Eqs. \eqref{eqnKet:FCB} runs over all values of the atomic
momentum  $\V{k}$. The integrands are the product of the momentum distribution at
$\V{k}'=\V{q}-\V{k}$ and of the diffraction function: The first accounts for the
effect of the photon recoil $\hbar\V{q}$ on the atomic distribution due to the
outcoupling process, while the diffraction function imposes energy conservation
in the scattering process. Equations~(\ref{eqnKet:FCB}a) and
~(\ref{eqnKet:FCB}b) show thus  that the contributions to the flux come from the
total number of condensed and noncondensed atoms, respectively, which fulfill
energy and momentum conservation of the scattering process.\\

We remark that according to Eq. (\ref{eqn:MLequiv}), the flux evaluated in Eqs.~(\ref{eqnKet:FCB}) gives also the corresponding component of the atomic flux. The latter agrees with the corresponding expressions derived in Ref. \cite{LuxatOutcoupling} for the atomic flux outcoupled from a single BEC by classical fields. 

\subsection{Interference contribution}

In order to determine the interference contribution to the photon flux, $F_I(t)$ in Eq. \eqref{F:I}, one needs the explicit form of the correlation functions $G_{LR}({\bf r},t;{\bf r'},t')$, $G_{RL}({\bf r},t;{\bf r'},t')$. For the considered setup, however, one can already make general statements. In absence of initial correlations, in fact, they are the product of the mean value of the field operators in each well, and thus take the form
\begin{eqnarray}
G_{LR}({\bf r},t;{\bf r'},t')&=&{\rm e}^{{\rm i}(\mu_Lt-\mu_Rt')/\hbar}\langle
\psi_L^\dagger({\bf r},t)\rangle  \langle\psi_R({\bf r'},t') \rangle\nonumber\\
&=&{\rm e}^{{\rm i}(\mu_Lt-\mu_Rt')/\hbar}{\rm e}^{-{\rm i}\varphi_L} f_L({\bf r}) \langle\psi_R({\bf
r'},t') \rangle  \nonumber \\ \label{G:LR}
\end{eqnarray}
while $G_{RL}({\bf r},t;{\bf r'},t')=[G_{LR}({\bf r'},t';{\bf r},t)]^*$. 

From Eq. (\ref{G:LR}) one observes that these correlation functions are
proportional to the mean value of atomic field operator $\langle\psi_R({\bf r},t) \rangle$. When this is zero, the interference contribution vanishes. Otherwise,
the amplitude of this term is proportional to the condensate fraction of the
quantum gas in the right well. We remark that the noncondensed atoms in the left
well do not contribute to the signal, since (i) there are no initial
correlations between the atoms in the left and right wells, and (ii) the mean
value $\langle\delta \psi_L({\bf r},t)\rangle=0$. 

In the shape of Eq. (\ref{G:LR}) the analogy with homodyne detection, as it is performed with light fields, can be drawn: the condensate in the left well plays the role of the local oscillator \cite{Homodyne}. Identifying the analon of the phase of the local oscillator is however a more delicate issue, that deserves some more analysis. For this
purpose, we first assume that the mean value $\langle\psi_R({\bf r},t) \rangle$
can be evaluated within a mean-field approach, such that  $\langle\psi_R({\bf
r},t) \rangle=f_R({\bf r}) {\rm e}^{{\rm i}\varphi_R}$, with $f_R({\bf r})$
real-valued function and $\varphi_R$ real constant. The integral in Eq.~(\ref{F:I}) can then be cast in the form
\begin{widetext}
\begin{eqnarray}
F_I(t)=2\pi \Gamma {\rm Re}\int \frac{{\rm d}\V{k}}{(2\pi)^3}  \left[{\rm e}^{{\rm
i}(\delta \mu t-\varphi_{LR})}\tilde f_L(\V{k})^* \tilde f_R(\V{k})\tilde
\delta^t(\Omega-\omega_{\V{k}+\V{q}})+{\rm e}^{-{\rm i}(\delta \mu
t-\varphi_{LR})}\tilde f_R(\V{k})^* \tilde
f_L(\V{k})\tilde\delta^t(\Omega-\delta\mu-\omega_{\V{k}+\V{q}})\right]\,,
\label{F:I:2}\end{eqnarray}\end{widetext}
with 
\begin{equation}\delta\mu=(\mu_L-\mu_R)/\hbar\,,
\end{equation}
 and 
\begin{equation}
\label{eqn:varphid_Def}
 \varphi_{LR} = \varphi_L-\varphi_R\,. 
\end{equation}
Equation \eqref{F:I:2} evidentiates that the interference contribution is an
oscillating signal. The phase of the oscillation, however, depends on time and
oscillates with a frequency determined by the difference $\delta\mu$ between the chemical
potentials, while the phase offset is determined by the relative phase between
the two quasi-condensate. Since the two quantum gases are independent, this
relative phase will be defined only for a single experimental run \cite{Pitaevski99,Janne,Demler_Lectures,Hadzibabic,Schmiedmayer,Iazzi}.

\subsection{Discussion}
\label{sec:phase}

\subsubsection{Which-way information and quantum erasers}

Some remarks on the oscillating behaviour of the interference term are now in order. In first place, for the example considered in Eq. (\ref{F:I:2}) the photon flux oscillates in time with frequency $\delta\mu$. This oscillation is observed, however, only when one chooses a time-window at the detector $\Delta t$ such that $\delta \mu\Delta t \ll 1$. Indeed, the photon flux is composed by two components: one is constituted by the atoms extracted from the left condensate, which interfere with the atoms extracted from the right trap during the time interval $\Delta t$ (first term on the RHS of Eq. \eqref{F:I:2}). The second component is constituted by the atoms extracted from the right trap, which interfere with the atoms extracted from the left condensate during the time interval $\Delta t$ (second term on the RHS of Eq. \eqref{F:I:2}). This interference is analogous to the interference between two lasers at different frequencies: interference fringes are observed provided that the time window of the detector is sufficiently small so not to resolve the detuning between the lasers \cite{Mandel,Paul}. As in the case of two independent lasers, there is no well defined phase offset: the relative phase is defined only for a single experimental run, while the average over a statistically significant number of runs gives no interference pattern \cite{Pitaevski99,Janne,Demler_Lectures,Schmiedmayer}.

Differing from the situation of two laser beams  \cite{Mandel}, however, photon scattering here creates correlations among the scattered photon, the corresponding outcoupled atom, and the quantum gas. This correlation is a form of a which-way information, that in general washes out any interference in the photon flux \cite{Englert_WhichWay} and can be considered as a form of photon-atoms entanglement \cite{Monroe}. Oscillations in the photon flux, and hence interference, can be recovered if (i) the beam of atoms outcoupled from one well overlap spatially with the wave function of the atoms in the second trap and (ii) if there are either initial correlations between the atoms in the two wells or if they both possess a non-vanishing condensate fraction.

Condition (i) corresponds to a quantum eraser \cite{scully_QuantumEraser}: when it is fulfilled, in fact, the outcoupled atom is no more entangled with the scattered photon. We will see in the following that it can be fulfilled for certain geometries and after a certain time, corresponding to the time needed for the outcoupled atoms to travel to the second well.  

Condition (i) alone is however not sufficient. The photon is not only correlated with the outcoupled atom, but also with the scattering system. Therefore, only when either the atoms in left and right system are correlated (via e.g. tunneling events which may wash out the which-way information \cite{Priscilla}) or when these correlations are partly classical, as it occurs when there is a finite condensate fraction in both systems, then interference can be reestablished. This is essentially condition (ii). 

We remark that oscillations of the photon flux as a function of time have been observed in this setup in Refs.~\cite{Saba_DetPhase,Shin_OptWeakLinkBEC}. Here, they have been discussed in terms of Josephson Junction, taking into account the possibility that the outcoupled atoms from one condensate can be then retrapped in the second well by means of an inverse Raman process. The discussion we just reported provides a quantum optical interpretation of the phenomenon, that is valid for short times, when the quantum state of the scattering atoms is not substantially perturbed by the scattering process. 

\subsubsection{Onset of the time-dependent oscillations}

We now analyse the onset of oscillations in time. As we mentioned, the photon flux starts oscillating after a finite time has elapsed from the beginning of the experiment.~\cite{Saba_DetPhase,Shin_OptWeakLinkBEC}. This can be also seen in our theory, by performing the integral over ${\bf k}$ in Eq. (\ref{F:I:2}). In App.~\ref{app:Int} we show that $F_I$ can be recast in the form $F_I(t)=F_{L\rightarrow R}(t) +F_{R\rightarrow L}(t) $, with 
\begin{subequations}\label{eqn:FCInt_posText}
\begin{eqnarray}
F_{L\rightarrow R}(t) &\approx &\Gamma\mathrm{Re}\,{\rm e}^{{\rm i} (\delta \mu
t-\varphi_{LR})
}\int_{0}^{t}{\rm d}\tau{\rm e}^{{\rm i}  (\omega_{\V{q}}-\Omega )\tau }
h_{LR}(\V{d},\tau) \, \\
F_{R\rightarrow L}(t) &\approx &\Gamma\mathrm{Re}\,{\rm e}^{-{\rm i} (\delta \mu
t-\varphi_{LR})
}\int_{0}^{t}{\rm d}\tau {\rm e}^{{\rm i}(\omega _{\V{q}}-\Omega +\delta
\mu)\tau }
 h_{RL}(-\V{d},\tau) \, . \nonumber \\
\end{eqnarray}%
\end{subequations}
Here,
\begin{equation}\label{eqn:HLR}
h_{jl}(\V{d},\tau)=  \int {\rm d}
\V{r} f_j\left(\V{r}+\V{d}-v_\V{q} \tau \right )f_l(\V{r}) 
\end{equation}
is the overlap between the left and the right condensate, with one being
displaced by the amount $v_\V{q}\tau-\V{d}$, with the recoil velocity 
\begin{equation}
\label{recoil:v}
v_\V{q}=\frac{\hbar \V{q}}{m}\,,
\end{equation}
that is acquired by the outcoupled atom by scattering the photon. This overlap
vanishes at time $\tau=0$, i.e., when the outcoupling lasers are switched on
(recall that the two clouds initially do not overlap). We note that
the overlap integral is zero at all times if $\V{d}$ and $\V{q}$ are
orthogonal, while it may become maximum after a certain time, when $\V{d}$ and $\V{q}$ are parallel, say, pointing along the positive $x$ axis. In this case, the component $F_{L\rightarrow R}(t)$ may not vanish and can be interpreted as
the contribution to the interference flux $F_{I}(t)$ from the outcoupled atoms
that propagate from left to right (for the term $F_{R\rightarrow L}(t)$ this is
just opposite). Interference will set in, provided that a sufficiently long time has elapsed to warrant overlap. This corresponds to times $t>t_c$ with $$t_c\sim \frac{d_{\rm eff}}{v_q}\,,$$ where $d_{\rm eff}=d-(\xi_L+\xi_R)/2$ is the effective distance of the two systems taking into account their width $\xi_j$. In terms of an interference experiment, for times $t>t_c$ the which-way information has been erased and oscillations in the photon flux can be observed. 

In interferometric setups the amount of visibility and which-way information are related by an inequality \cite{Englert_WhichWay}. It is therefore useful to determine a visibility of the oscillating signal, as it contains information on the properties of the scattering systems. A visibility can be defined for sufficiently long times, averaging over several oscillating periods after the instant $t_c$, and will be proportional to the amplitude of the oscillations of the photon flux, hence to the product of the condensate fractions of both systems.
We remark, once again, that oscillations can be observed only in a single experimental run, while they will disappear after performing an ensemble average. Therefore, this behaviour can only be measured in systems, whose properties are not deeply modified by the outcoupling of atoms. When this is not verified, the presence of a condensed fraction in the right system can be revealed by performing an
ensemble average over a sufficient large number of experiments, in which the
signal is monitored for sufficiently short times, warranting that the properties
of the quantum gas have not been significantly modified, and taking the statistical distribution of the intensity of the photon flux at a given instant of time. An amplitude can be extracted from the width of the distribution by taking into account the finite width of the diffraction function.

\section{Homodyne detection of a quantum gas}
\label{Sec:3}

The theory presented so far will be now applied to some specific examples. The
main idea is to use the setup in Fig. \ref{fig:HomodyneSetup} in order to
determine the mean value of the field operator of a quantum gas, using a 
Bose-Einstein condensate at known temperature as reference system. Such setup is
a matter-wave analogon of homodyne detection in quantum optics. The individual elements can be so identified:  the BEC acts as a local oscillator, the outcoupling procedure as beam splitter, the relative phase can be varied by changing the interwell distance. The information on the atomic gas is
carried by both scattered photons and outcoupled atoms: homodyne detection of
the scattered field, hence, allows one in principle to determine the mean value of the quadrature of the atomic field operator.

\subsection{Interference between two Bose-Einstein Condensates}

We first discuss the case in which two BEC are trapped in the left and right
well, respectively, and are both illuminated by the pump and probe beams. The
outcoupled atoms  from both condensates propagate along the direction determined by the distance between the minima of both wells, and overlap after the time $t_c$ has elapsed. The scattered photons are revealed at a detector in the far-field. This setup has been realized in the experiment reported in Refs. \cite{Saba_DetPhase,Shin_OptWeakLinkBEC}, where time-dependent oscillations in the atom and photon flux were measured. Here, we apply the theoretical model developed so far and find it reproduces qualitatively the results of these experiments. Moreover, we discuss the results in the light of question addressed in the present work.

We assume two BEC with equal number of atoms $N_C$ and temperature $T$, that are
confined either in the left or right well. The wells are described by the
potential
\begin{equation}\label{eqn:trappot}
V_{\{j=L,R\}}(\V{r})=V(\V{r}-\V{r}_j)+\delta_{j,L} \Delta V
\end{equation} 
with $V(\V{r})= \frac{1}{2}m(\omega _{x}^{2}x^{2}+\omega
_{y}^{2}y^{2}+\omega _{z}^{2}z^{2})$,  and $\Delta V$ denotes the constant
offset between the two traps.  For simplicity we take that both BEC are at zero
temperature, $T=0$, and the atoms weakly interact, such that the contribution of
the noncondensed atoms to the photon flux is small and can be neglected.  In
this limit the chemical potential of both condensates is equal and given by
$\mu(0)$. With definition (\ref{eqn:trappot}) then $\mu_R=\mu(0)$ and
$\mu_L=\mu(0)+\Delta V$. 

The photon flux is evaluated using the Thomas-Fermi approximation for the
condensate wave functions \cite{Stringari_Book},
\begin{equation}\label{eqn:ThomasFermi}
 f_j({\bf r})=[(\mu(0)-V(\V{r}-\V{r}_j))/g]^{1/2}\, ,
\end{equation}
where $\mu(0)=\left(15N_C a_s/\bar{a}   \right)^{2/5}\hbar \bar{\omega}/2$ is
the chemical potential at zero temperature and
$\bar{a}=\sqrt{\hbar/(m\bar{\omega})}$ is the size of the ground state of a
harmonic oscillator with frequency
$\bar{\omega}=(\omega_x\omega_y\omega_z)^{1/3}$. The condensate macroscopic wave
function has size $r_{\{\ell=x,y,z\}}^{(0)}=\sqrt{2\mu(0)/(m\omega_\ell^2)}$. Its Fourier
transform reads $\tilde{f}_L({\bf k})={\rm e}^{{\rm i}{\bf k}\cdot {\bf
d}/2}\tilde{f}_0({\bf k})$ (for the right condensate
$\tilde{f}_R({\bf k})={\rm e}^{-{\rm i}{\bf k}\cdot {\bf d}/2}\tilde{f}_0({\bf
k})$), where $\tilde{f}_0({\bf k})$ is the Fourier transform of the macroscopic
wave function centered at the origin and is real valued. In particular, 
\begin{equation}
\tilde{f}_0({\bf k})=\kappa_0\frac{|J_2(p_0)|}{p_0^2}\,,
\end{equation}
where $\kappa_0=\sqrt{15 \pi^3 N_C r^{(0)}_x r^{(0)}_y r^{(0)}_z/2}$ is a scalar
and $J_2(p_0)$ the Bessel function of second order of the variable $p_0$ defined
as $p_0^2=k_x^2r_x ^{(0)2}+k_y^2r_y^{(0)2}+k_z^2r_z^{(0)2}$, see \cite{Stringari_Book}.  The integral for
the interference contribution to the photon flux in Eq. (\ref{F:I:2}) can be
then cast in the form
\begin{widetext}
\begin{equation}
\label{eqn:FluxesMomentumKetterle}
F_I(t)=2\pi \kappa_0^2\Gamma\int \frac{{\rm d}{\bf
k}}{(2\pi)^3}\frac{|J_2(p_0)|^2}{p_0^4}
{\rm Re}\left[{\rm e}^{{\rm i}(\delta\mu t-{\bf k}\cdot {\bf
d}-\varphi_{LR})}\tilde\delta^t(\Omega-\omega_{{\bf k}+{\bf q}})
+{\rm e}^{-{\rm i}(\delta\mu t-{\bf k}\cdot {\bf
d}-\varphi_{LR})}\tilde\delta^t(\Omega-\delta\mu-\omega_{{\bf k}+{\bf q}})
\right]\,.
\end{equation}
This equation can be simplified, taking that both $\V{d}$ and $\V{q}$ point
along the positive $x$-direction. Following the derivation reported in App.~\ref{app:Ketterle}, we find that the total flux can be approximated by the expression
\begin{equation} \label{eqn:Ftpostot}
F(t) \approx 2 \pi \Gamma N_C K_m
\left [ 1+\mathcal V_0
\cos \left (\delta \mu t-\varphi_{LR}+(\omega_{\V{q}}-\Omega )\frac{d}{v_{q}}
\right )\Theta \left(t-\frac{d-2r_x}{v_q}\right) \right ] \, ,
\end{equation}
\end{widetext}
where $$K_m=(K(\omega_\V{q}-\Omega)+K(\omega_\V{q}-\Omega+\delta \mu))/2$$ and $K(x)
= \sqrt{1/(2 \pi \sigma^2 ) } \exp[-x^2/(2 \sigma^2)] $ is a Gaussian of width
$\sigma^2=2.5(v_\V{q}/r_x)^2$.  Equation \eqref{eqn:Ftpostot} shows that the photon
flux starts oscillating for times $t>t_c$, with $t_c=(d-2r_x)/v_\V{q}$, namely, when the
outcoupled atoms from one condensate have reached the second one. The time-dependent oscillations have frequency $\delta
\mu$ and amplitude
\begin{equation}
\mathcal
V_0=2\frac{K(\omega_\V{q}-\Omega)}{K(\omega_\V{q}-\Omega)+K(\omega_\V{q}-\Omega+\delta
\mu)}\,.
\end{equation}
The expression in Eq. (\ref{eqn:Ftpostot}) has been obtained neglecting the momentum dependence of the trapped clouds (this approximation is verified in  App.~\ref{app:Ketterle}) and the contribution of the noncondensed
fraction. We note that, although oscillations as a function of time are observed
provided that $\delta\mu\neq 0$, nevertheless their visibility (which corresponds to $\mathcal V_0$) is always smaller than unity, $\mathcal V_0<1$ for $\delta\mu\neq 0$.  This can be understood
considering that due to energy conservation in the scattering process, a
scattered photon carries partial information about from which cloud it was
scattered. 

We now turn our attention to the phase offset, characterising the oscillations
of the photon flux as a function of time in Eq.~(\ref{eqn:Ftpostot}). This phase
offset is here given by the quantity $\Theta-\varphi_{LR}$, where $\varphi_{LR}$
is the relative phase between the two BEC, which can take any value (we refer to
the discussion in Sec. \ref{sec:phase}), and by the quantity
\begin{equation} \label{eqn:PhaseOffset1}
\Theta =\frac{d}{v_{q}}(\omega _{\V{q}} -\Omega+\delta\mu)
\, ,
\end{equation}%
which is the phase the outcoupled atoms accumulate when travelling from left to the right well
(The corresponding phase offset for the outcoupled atoms travelling from the right to the left well is 
given in App.~\ref{app:Ketterle}). This phase can be varied by either
tuning the laser frequency, and thus $\Omega$, or changing the distance between
the minima of the two wells.  The phase offset in Eq. \eqref{eqn:PhaseOffset1}
agrees with the one reported in Ref. \cite{Shin_OptWeakLinkBEC}, that was
derived from a phenomenological model.   In  Ref. ~\cite{Shin_OptWeakLinkBEC}, in
particular, the rate of change of the difference in phase offset between atoms
outcoupled to the left and atoms outcoupled to the right  was measured  as a function of the
Bragg-frequency $\Omega$, and found in agreement with the
prediction of Eq.~(\ref{eqn:PhaseOffset}).

We also note that our theory predicts correctly the oscillation period observed
in the experiment, and also accounts for the fact why the interference current
needs some time to build up. The linear response treatment we apply, however,
does not predict any decrease in the visibility of the interference pattern with
time, in marked contrast with the experimental results. Possible reasons for the
decrease in visibility in the experimental data are depletion of the condensate
and heating of the sample, which could be due to spontaneous Rayleigh scattering
events \cite{Saba_DetPhase}. These effects are not taken into account in our model, but
could be introduced by means of quantum Langevin equations, using a
formalism similar to the one developed in Ref. \cite{Larson_NJP}.

\begin{figure}[htp] \centering
\includegraphics[width=.44\textwidth]{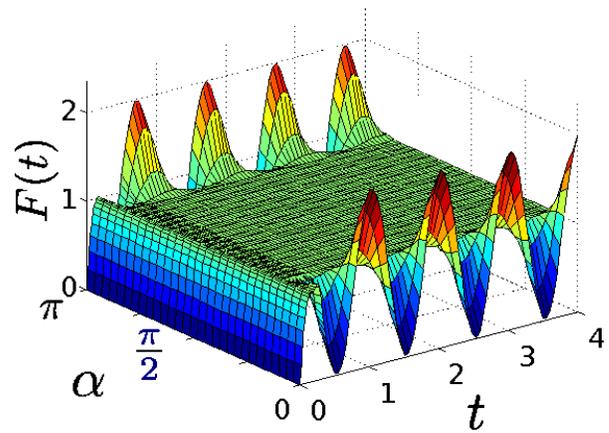}
\caption{\label{fig:FluxesMomentum}  Photon flux $F(t)$ (in units
of the background contribution $F_B(t)$) as a function of time (in units of $\delta \mu/2\pi$) and of the angle
$\alpha$ at which the atoms are emitted. The photon flux is computed by
numerically integrating Eq.~(\ref{eqn:FluxesMomentumKetterle}). 
The condensates are composed by $N=10^6$ sodium atoms in a spherical harmomic trap of frequency $\omega=325 {\rm Hz}$. The scattering length is $a_s=55 a_0$ with $a_0$ Bohr radius. The other parameters are
$v_\V{q}=6\frac{\rm cm}{\rm s}$, $\delta \mu_0 = 2\pi 10^3 {\rm Hz}$, $d=5 r_x$,
$\Omega=\omega_\V{q}$, $\alpha=0$ and $\varphi_{LR}=0$.}
\end{figure}

We finally discuss the dependence of the photon flux on time and on the angle of
emission of the outcoupled atoms, thus for geometries where the
direction of emission $\V{q}$ of the outcoupled atoms forms an angle $\alpha$
with the vector ${\bf d}$. For simplicity we assume that both vectors lie in the
$x-y$ plane. The calculation is performed by numerically integrating
Eq.~(\ref{eqn:FluxesMomentumKetterle}). 

Figure~\ref{fig:FluxesMomentum} displays the photon flux $F(t)$ as a function of
time and of the emission angle $\alpha$. Oscillations as a function of time are
observed for values of the angle about $\alpha= 0,\pi$, corresponding to the
atoms propagating in the direction parallel to $\pm \V{d}$. The period of the
oscillation is $2\pi/\delta\mu$, independent of the angle of emission. The
oscillations disappear for angles in the interval $\pi/8\lesssim\alpha\lesssim
7\pi/8$: the photon flux is here solely given by the background contribution
$F_B(t)$. For these angles, in fact, at all times there is
no spatial overlap between the wavefunction of the outcoupled atoms and the
wavefunction of the trapped atoms in the other well. Indeed, from a simple
geometric argument one finds that the overlap vanishes for angle $\alpha>\arctan
\frac{2 r_x}{d}\approx \pi/8$ and $\alpha<\pi-\pi/8$. This implies that the setup must be so
constructed, that the atoms outcoupled from one well could in principle be
transferred, by a Raman process, into the second well. This property is the key
element on which the analogy to a Josephson Junction has been drawn
\cite{Shin_OptWeakLinkBEC}. It is also basically the way in which a quantum eraser is realised in this setup.
We refer the reader to the discussion on  the properties of this quantum interference process in Sec. \ref{sec:phase}.

\subsection{Thermometry of a Bose-Einstein condensate}

We now show how the setup of Ref. \cite{Saba_DetPhase} could be used in order to
determine the temperature of a Bose-Einstein condensate, using as reference a
second BEC at known temperature. For this purpose, we make the same assumptions
as in the previous section, with the difference that while the left gas is a BEC
at temperature $T=0$, the gas in the right well is a BEC at temperature $T$ to
be determined. We further assume that the Thomas-Fermi approximation can be
performed also for the right condensate. Hence, the chemical potential of
the second condensate can be written as $\mu(T)=\mu(0)(1-T^3/T_c^3)^{2/5}$, with
$T_c$ the critical temperature for the noninteracting gas in a harmonic trap, while the size of the macroscopic wave function of the right condensate
scales with $r_\ell^{(T)}=r_\ell^{(0)}(1-T^3/T_c^3)^{1/5}$, see \cite{Stringari_Book}. Using these
relations the integral for the interference contribution to the photon flux can
be cast in the form 
\begin{widetext}
\begin{eqnarray}
\label{F:I:T}
F_I(t)
&=&2\pi\Gamma\kappa_0^2n_C(T) \int
\frac{{\rm d}{\bf k}}{(2\pi)^3}\frac{|J_2(p_0)J_2(p_T)|}{p_T^4}\nonumber \\ & &
\times {\rm Re}\left[{\rm e}^{{\rm i}(\delta\mu(T) t-{\bf k}\cdot {\bf
d}-\varphi_{LR})}\tilde\delta^t(\Omega-\omega_{{\bf k}+{\bf q}})
+{\rm e}^{-{\rm i}(\delta\mu(T) t-{\bf k}\cdot {\bf
d}-\varphi_{LR})}\tilde\delta^t(\Omega-\delta\mu(T)-\omega_{{\bf k}+{\bf q}})
\right]\,,
\end{eqnarray}
\end{widetext}
with $p_T=(1-T^3/T_c^3)^{1/5}p_0$.  The integral depends on the temperature both
through a scaling factor as well as the function $J_2(p_T)/p_T^4$ in the integrand,
while the phases depend on the temperature of the second BEC via the chemical
potential of the right BEC, which enters in the quantity
\begin{equation}
\delta\mu(T)=\delta\mu(0)-\mathcal F(N,T)/\hbar
\end{equation}
with $\mathcal F(N,T)=\mu(T)-\mu(0)$.  We note that $F_I(t)$ oscillates as a
function of time with both frequency and amplitude which depend on the
temperature of the second condensate. For $T\ll T_c$, in particular, one finds that
$F_I(t)\propto \sqrt{n_C(T)}$, where  $n_C(T)=N_C(T)/N=1-\left(T/T_c\right)^3$ is
the condensate fraction in the right trap. 

\begin{figure}[htp] \centering
\subfigure[]{
\includegraphics[width=.45\textwidth]{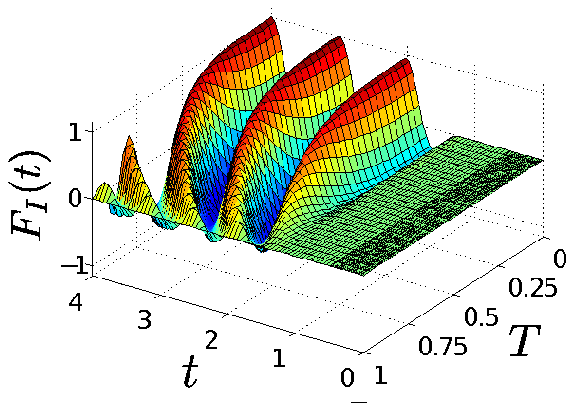} 
}
\subfigure[]{
\includegraphics[width=.45\textwidth]{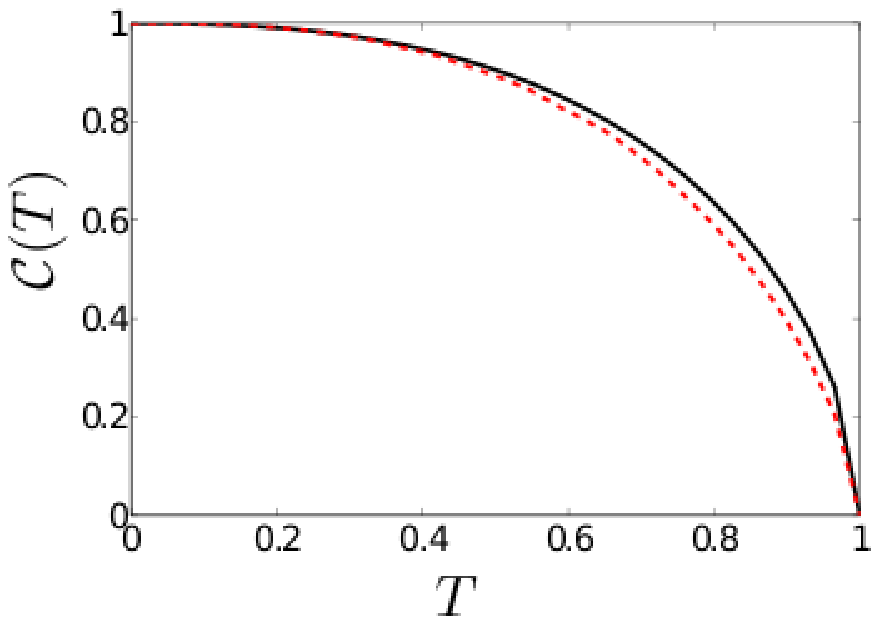} 
}
\caption{\label{fig:FLRTemperature} (color online) (a) Interference term of the Photon flux, $F_I(t)$, Eq.~\eqref{F:I:T}, (in units of the background current obtained when both condensates are at zero temperature $F_B(T=0)$), as a function of time (in units of $2\pi/\delta \mu$) 
and temperature $T$ (in units of critical temperature $T_c^i$).  (b) Amplitude $ \mathcal C (T) $,
Eq.~(\ref{eqn:Atr}) (in units of $ \mathcal C (0) $) as a function of the temperature $T$
(in units of $T_c^i$). The red dashed line corresponds to the squared root of the condensate fraction in the right BEC, $\sqrt{n_C(T)}$. The other parameters are as in Fig.~\ref{fig:FluxesMomentum}. }
\end{figure}

Figure~(\ref{fig:FLRTemperature}a) displays $ F_I(t)$ as  function of time and
temperature $T$ of the right condensate, when the trap is spherical. The oscillations are visible for times $t>t_c$. The oscillation frequency depends on $T$, as one can clearly
observe from the figure \cite{CondDepletion}. The
dependence of the amplitude of the oscillation on the temperature is visible in Fig.~(\ref{fig:FLRTemperature}a) and is singled out in
Fig.~(\ref{fig:FLRTemperature}b), where the amplitude  
\begin{equation} \label{eqn:Atr}
  \mathcal C (T) = [{\rm max} (F_I(t)) -{\rm min} (F_I(t))] \, ,
 \end{equation}
evaluated at times $t>t_c$, is displayed as a function of $T$ in units of $C(0)$. The red
dashed curve represents the squared root of the condensate fraction, $\sqrt{n_C(T)}$, and is reported for comparison.
Function $\mathcal C(T)$ decreases monotonically as the temperature increases
from $T=0$, and vanishes at the critical temperature $T=T_C^i$ (which takes into
account the effect of the interactions and is such that $n_C(T_c^i)=0$) \cite{footcrittemp}. 

\subsection{Monitoring the Mott-Insulator/Superfluid transition}

\begin{figure}[htp] \centering
\subfigure[]{
\includegraphics[width=.47\textwidth]{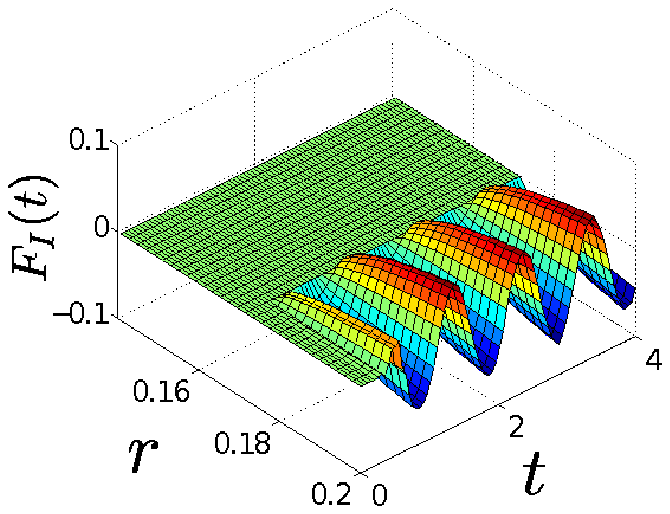} 
}
\subfigure[]{
\includegraphics[width=.41\textwidth]{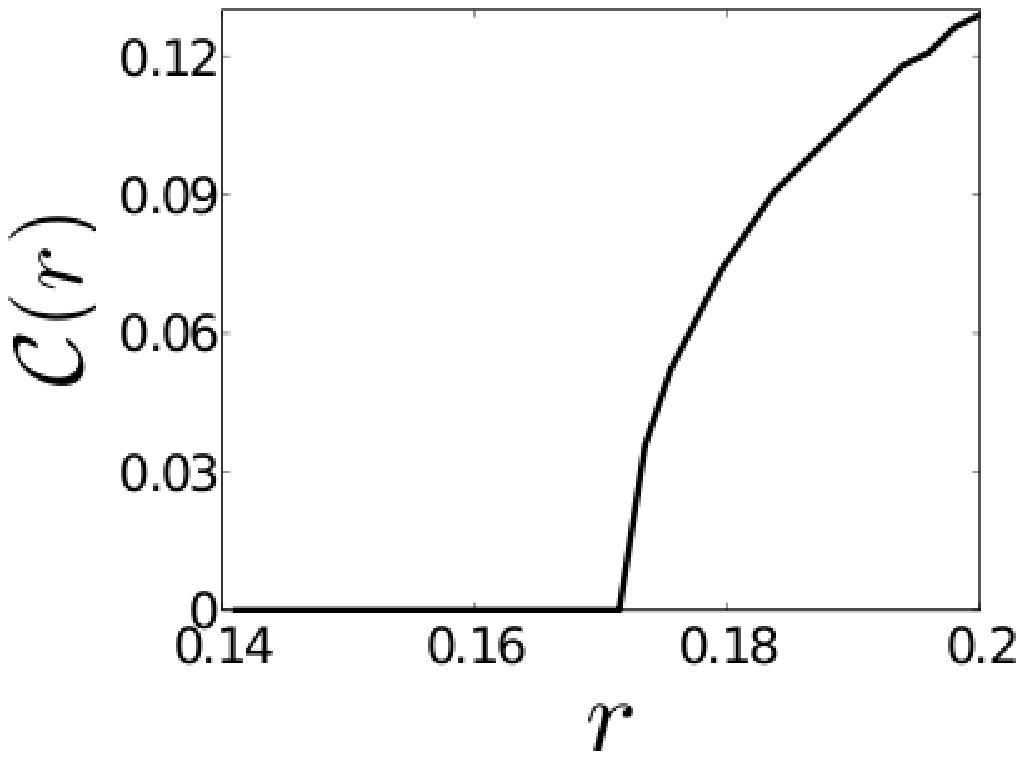} 
}
\caption{\label{fig:FLRandFa} (color online) (a) Interference contribution $F_I(t)$ 
as a function of time (in units of $2\pi/\delta \mu$) and of the parameter $r=zJ/U$. $F_I(t)$ is given in units of  $\Gamma M $, where $M=50\times 50$ is the number of lattice sites. (b) Amplitude $\mathcal C(r)$ (in units of  $\mathcal C(0)$) as a function of $r$. The parameters are $V_0=10\, \hbar \omega_R$,
$x_0=0.13 d_0$, $\delta \mu = 0.2 \omega_R$, $d=20d_0$, $|{\bf q}|=2\pi/d_0$,
$\Omega-\omega_\V{q}=\omega_R$, with $\omega_R=\hbar\pi^2/(2 md_0^2)$, and $\varphi_{LR}=0$. Moreover, $\mu_R/U=\sqrt{2}-1$ and $n_R=\expct{b}^2 |_{r=0.2}/d_0^2\approx 0.2/d_0^2$.} 
\end{figure}

We now assume that the quantum system in the right trap is a two-dimensional
optical lattice, where the parameters can be tuned so that the ultracold atomic
gas is either in a Mott-insulator or superfluid state, while the left system is
a BEC at $T=0$. Within a mean-field treatment, in the superfluid phase a
nonvanishing superfluid order parameter would give rise to a time-oscillating
behaviour of the photon flux \cite{Sachdev}. Before we start, we recall that the
Mott-insulator/superfluid transition in two dimensions has been experimentally characterised in
Ref. \cite{Spielman}, and the condensate fraction has been measured in Ref. \cite{ CondFracPhillips}, while Bragg scattering by this system was recently reported in Ref. \cite{LightScattMottBloch}. 

Here, we argue that the photon flux may allow one to detect the superfluid
order parameter, and thus monitor the quantum state of the system across the
phase transition. We note, however, that a measurement based on outcoupling atoms from the
optical lattice will change the phase of the atoms, if they have been prepared in the Mott-insulator phase. For this specific situation, indeed, the approximation we make by treating photon scattering as a weak
perturbation of the many-body state remains valid as long as the scattering is
performed over a time-scale in which the creation of the local defect does not
affect the rest of the system. This time scale will be inversely proportional to
the sound velocity with which a perturbation propagates in the corresponding
superfluid state. This restriction shows that this measurement procedure will
not provide a reliable estimate of the superfluid order parameter close to the
phase transition point.

Keeping this in mind, let us now assume that the two atomic gases are at $T=0$
and tightly confined in the $x$-direction by a harmonic potential, 
so that the motional degrees of freedom in
the $x$-direction are frozen out and the systems are essentially two
dimensional. The left potential is assumed to be a highly-anisotropic harmonic
trap and right potential is a two-dimensional optical lattice. They read
\begin{eqnarray}
&&V_L(\V{r})= \frac{1}{2} m \omega_x^2 x^2 +\frac{1}{2} m \omega_\perp^2
(y^2+z^2)\, ,\\
&&V_R(\V{r})= \frac{1}{2} m \omega_x^2 x^2 + V_0 \left ( \sin^2 \frac{\pi
y}{d_0} + \sin^2 \frac{\pi z}{d_0} \right ) \label{eqn2Doptpot} \, .
\end{eqnarray}
with $\omega_\perp \ll \omega_x$, while $V_0$ is the height of the optical
lattice potential and $d_0$ the spatial periodicity in the transverse
directions. 

Assuming that the left quantum gas forms a weakly-interacting Bose-Einstein
condensate at $T=0$, and that $\hbar\omega_x\gg \mu_L(0)$, with $\mu_L(0)$ the
chemical potential of the left condensate, then the macroscopic wave function
reads $\Phi_L(\V{r}) = f_0(x) \sqrt{n_L} e^{{\rm i} \varphi_L}$, with $n_L$ is
the planar density of condensed atoms in the left trap (which is here assumed to
be constant, under the condition that the laser beams illuminates the center of
the condensate) and $f_0(x)=\frac{1}{\sqrt{x_0\sqrt{\pi}}} e^{-x^2/(2x_0^2)}$ is
the motional ground state of the harmonic oscillator in the $x$-direction. 
The wave function for the right quantum gas is here given assuming the
tight-binding limit. Using the Wannier decomposition, in the single-band limit one finds 
\begin{equation} \label{eqn:Phiexpand}
 \psi_R(\V{r})= f_0(x) \sum_{l,m} w_l(y)w_m(z) b_{l,m} \, ,
\end{equation}
where $w_l(\xi)$ is the Wannier function centered at $\xi_l=ld_0$ ($\xi=y,z$)
and $b_{l,m}$ annihilates an atom at the lattice site with $y=ld_0$ and
$z=md_0$. 
The corresponding Bose-Hubbard model reads~\cite{Jaksch_BH}
\begin{eqnarray}
H_{BH}^{(R)}&=&-J\sum_{l,m}\left(b_{l,m}^{\dagger}(b_{l-1,m}+b_{l,m-1})+{\rm
H.c.}\right)\nonumber\\
 & &+U\sum_{l,m}n_{l,m}(n_{l,m}-1)-\mu_R\sum_{l,m}n_{l,m}
\end{eqnarray}
with $n_{l,m}=b_{l,m}^{\dagger}b_{l,m}$ the number of atoms at site $(l,m)$, $J$
the hopping rate, $U$ the onsite interaction, and $\mu_R$ the chemical potential
for the right system. We note that the two-dimensional limit is consistent when
$\hbar\omega_x\gg\mu_R$. In the superfluid phase the expectation value of
operator $b_{lm}$ over the lattice ground state does not vanish and is a
constant, $\langle b_{l,m}\rangle=\langle b\rangle$, and the wavefunction for
the superfluid fraction reads $\Phi_R(\V{r}) = f_0(x) \expct{b} \sum_{l,m}
w_l(y)w_m(z)$.

The photon flux, and in particular the interference contribution, can now be
evaluated using these quantities. For simplicity, in the following we assume
that the geometry of the laser is such that $\V{q}$ is parallel to the vector
$\V{d}$. 
An analytical expression can be determined using the Gaussian ansatz for the Wannier functions \cite{Kohn}
and shows that $F_I(t)\propto \expct{b}$ and is a convolution of signals
oscillating with phase $\varphi_{LR}+k_x d-\delta \mu t$, weighted by the
occupation of the momentum $k_x$ fulfilling energy conservation.     

We evaluate $F_I(t)$ by numerically integrating Eq.~\eqref{F:I:2}. The
superfluid order parameter $\expct{b}$ is determined numerically within
the mean-field treatment of the Bose-Hubbard model~\cite{Sachdev}.
Figure~\ref{fig:FLRandFa}(a) displays the interference contribution $F_I(t)$ as a
function of time and of the ratio $r=zJ/U$ between hopping strength and the
onsite interaction, with $z$ the coordination number. This ratio is assumed
to be varied by changing the onsite interaction $U$, while the lattice depth
$V_0$ is kept constant. 
For small values of $r=zJ/U$, when the atoms in the right well are in the
Mott-insulator state, no interference signal is found. At a critical value
$(zJ/U)_c$ (unit filling) \cite{Wessel} the right system undergoes the quantum phase transition to the
superfluid state and the interference photon flux $F_I(t)$ starts to oscillate
in time with finite amplitude. We study the amplitude of the oscillations as a function of $r$ by
characterizing the quantity\begin{equation}
\mathcal C(r)={\rm max} (F_I(t)) -{\rm min} (F_I(t)) \, ,
\end{equation}
where the minimum and maximum of $F_I(t)$ are found over a time interval large
compared to the inverse oscillation frequency of $F_I(t)$, similarly to
Eq.~(\ref{eqn:Atr}). This quantity is displayed in Fig.~\ref{fig:FLRandFa}(b) as
a function of $r$. For $r>r_c$ the amplitude $\mathcal C(r)$ is nonvanishing and
increases proportional to the magnitude of the superfluid order parameter. 

\section{First-order correlation function}
\label{Sec:4}

Several methods have been proposed in the literature for determining the first-order correlation functions, that are based on time-of-flight techniques, see for instance Refs. \cite{Duan,Demler_Lectures,CarusottoPRL2007}. Measurements of the first-order correlation function have been performed on Bose-Einstein Condensates in Refs. \cite{SpatOrderBloch,LongRangeEsslinger2007}. 

In the following we will show how an extension of our previously considered
setup may allow one to determine the spatial first-order correlation function of a
quantum gas by the scattered photons.  
In this case, the lasers shall illuminate two spatially separated regions of a quantum gas confined in a {\it single well} potential. More specifically, the spatial dependence of the laser-atom interaction in Eq. \ref{eqnKet:V:0} will be characterized by the Gaussian envelope
$$|\gamma_j(\V{r})|=\gamma_0\exp\left(-(\V{r}-\V{r_j}\right)^2/\Delta\V{r}
^2)/(\sqrt{\pi}\Delta\V{r})^{1/2}\,,$$
with width $\Delta\V{r}\ll|{\bf d}|$ between the two regions (We remark that the excitation could be realized with subwavelength resolution \cite{Greiner_Science_2010,Bloch_Nature_2010,Gorshkov}). The corresponding photon flux then reads 
\begin{eqnarray}
\label{F:t:G:1}
F(t) &=& \Gamma \sum_{j,k=L,R}{\rm  Re}  \int_0^t {\rm d} t' \int{\rm
d}\V{r}\int{\rm d}\V{r'} 
\nonumber \\ && \times \gamma_j(\V{r})\gamma_k(\V{r'})f(\V{r},t;\V{r}',t')
G^{(1)}(\V{r},t;\V{r'},t')
\end{eqnarray}
where
\begin{equation}
\label{G:1:0}
G^{(1)}(\V{r},t;\V{r'},t')=\expct{\psi_1^\dagger(\V{r},t)\psi_1(\V{r'},t')} \,  
\end{equation}
is the first-order correlation function and $f(\V{r},t;\V{r}',t')$
is defined in Eq.~(\ref{fun:f}). Let the pump-probe excitation be a pulse of mean duration $t$ 
such that $\tilde{\omega}t\gg 1$, but $\omega_{\alpha}t\ll 1$, with $\omega_{\alpha}$ the typical frequency characterising the excitation spectrum. In this limit we can approximate  $G^{(1)}(\V{r},t;\V{r'},t')\simeq G^{(1)}(\V{r},0;\V{r'},0)$ in Eq. (\ref{F:t:G:1}). For convenience, we denote by $G^{(1)}(\V{r};\V{r'})\equiv G^{(1)}(\V{r},0;\V{r'},0)$ the spatial correlation function. 
For $\Delta {\bf r}$ sufficiently small, so that it can be approximated with a $\delta$-like excitation, the photon flux can be recast in the form
\begin{eqnarray}
\label{F:t:G:3}
F(t) &\simeq& K \left(1+{\rm  Re} \left\{{\rm e}^{{\rm i}{\bf q}\cdot
{\bf d}}G^{(1)}({\bf d/2};-{\bf d/2})\right\}/n_0({\bf d/2})\right)\nonumber\\
\end{eqnarray}
with $K$ is a constant, determined by the details of the excitation scheme, and 
$n_0({\bf d/2})=G^{(1)}({\bf d/2};{\bf d/2})$ is the density at ${\bf r}=\pm{\bf
d/2}$ (assuming the system has reflection symmetry about ${\bf r}=0$). Therefore, the photon flux exhibits oscillations with a visibility which
is determined by the spatial first-order correlation function. By varying the scattering
wave vector ${\bf q}$ one would thus measure the first-order correlation
function as a function of ${\bf d}$. Realistic excitation schemes are of course characterized by finite spatial resolution $\Delta {\bf r}$, which results in averaging the correlation function in Eq. \eqref{F:t:G:3} over the finite size of the illuminated region and hence to a diminution of the contrast \cite{Priscilla}. 

This setup could be extended to measure time-dependent correlation function by applying a pair of laser pulses: Assuming that the photon scattered after the first pulse can interfere with the photon scattered after the second pulse, then the photon flux at the detector will exhibit oscillations whose amplitude is proportional to the first-order correlation function \eqref{G:1:0}. A possible realization could use a mirror placed in front of the quantum system, as realized in \cite{Eschner}.

\section{Conclusions}
\label{Sec:5}

In this work we have discussed a setup which allows one, to measure the condensate fraction and the first-order correlation function of a quantum gas by means of photo-detection. The photons are scattered by the quantum gas in a pump-probe type of excitation, such that the scattered photon is associated with an outcoupled atom, with which it is entangled. In addition, photon and atom are correlated with the quantum gas. This correlation is detected in the photon flux, provided that certain conditions are fulfilled, which we have identified and discussed. 

Our analysis is based on the impulse approximation \cite{Pitaevski99}, hence on neglecting the back action of the scattering process on the quantum gas. It is therefore valid for short time transients. From the point of view of collecting a sufficient statistics, hence, time-of-flight techniques are a more convenient tool. Nevertheless, one could consider to modify existing techniques, such as the one successfully demonstrated in Refs. \cite{Greiner_Science_2010,Bloch_Nature_2010}, in order to access to the same kind of information that the setup here discussed provides. 

An interesting outlook is to identify a setup, along the lines of the proposal  Ref. \cite{Polzik}, which permits one to perform a quantum-non-demolition measurement of any correlations function of of the external degrees of freedom of atomic gases, and more in general, which realizes quantum-state transfer between matter and light. This would open several interesting perspectives for quantum communications \cite{Chang}.

\acknowledgements

The authors acknowledge E. Demler, R. Fazio, J. Ruoteskovki, W. Schleich, and P.
Vignolo for stimulating discussions and helpful comments. This work was
supported by the European Commission (EMALI, MRTN-CT-2006-035369; Integrating
project AQUTE, STREP PICC), by the Spanish Ministerio de Ciencia y Innovaci\'on
(Consolider-Ingenio 2010 QOIT; FIS2007-66944; AI HU2007-0013, EUROQUAM
``CMMC''), and the German Research Foundation (DFG, MO1845).

\begin{appendix}

\section{} \label{app:Int}

We recast the interference term, Eq. (\ref{F:I}), in the form
$F_I=F_{L\rightarrow R}(t) +F_{R\rightarrow L}(t)$, with
\begin{subequations} \label{eqn:intF}
\begin{eqnarray} 
F_{L\rightarrow R}(t) &=&\Gamma\mathrm{Re} \,{\rm e}^{{\rm i} \delta \mu
t}\int_{0}^{t}{\rm d}t^{\prime }\int \frac{{\rm d}\V{k}}{(2 \pi)^3}
{\rm e}^{-{\rm i}(\Omega-\omega_{\V{k}}) (t-t^{\prime })}  \nonumber \\ & & \times
 \int {\rm d}\V{r}\,
{\rm d}\V{r}^{\prime} {\rm e}^{{\rm i}(\V{q}-\V{k}) \cdot ( \V{r}-\V{r}')} 
\Phi_L(\V{r}')^*\Phi_R(\V{r}) \, ,\\
F_{R\rightarrow L}(t) &=&\Gamma\mathrm{Re} \,{\rm e}^{-{\rm i} \delta \mu t}
\int_{0}^{t}{\rm d}t^{\prime }\int \frac{{\rm d} \V{k}}{(2 \pi)^3}
{\rm e}^{-{\rm i}(\Omega-\omega_{\V{k}}-\delta \mu) (t-t^{\prime })}  \nonumber \\ & &
\times \int {\rm d}\V{r}\,
{\rm d}\V{r}^{\prime} {\rm e}^{{\rm i}(\V{q}-\V{k}) \cdot ( \V{r}-\V{r}')}  \Phi_R(\V{r}')^*
\Phi_L(\V{r}) \, .\nonumber \\
\end{eqnarray}
\end{subequations}
We now show that the term $F_{L\rightarrow R}(t)$ ($F_{R\rightarrow
L}(t)$) is the contribution due to the atoms which are outcoupled and propagate from the
left to the right (right to the left). For this purpose we perform the $\V{k}$ integral
in Eq.~(\ref{eqn:intF}) and obtain
\begin{widetext}
\begin{eqnarray} \label{Hom:Propposa}
F_{L\rightarrow R}(t) =\Gamma\mathrm{Re}\,{\rm e}^{{\rm i} (\delta \mu
t-\varphi_{LR})
}\int_{0}^{t}{\rm d}\tau \left( \frac{{\rm i}
\,m}{2\pi
\hbar \tau }\right) ^{\frac{3}{2}}\int {\rm d}\V{r}\,\int {\rm d}\V{r}^{\prime}\, {\rm e}^{ {\rm i} ({\bf q}\cdot
(\V{r}-
\V{r}^{\prime })-\Omega \tau )} \exp \left[ -\frac{{\rm i} }{\hbar }\frac{m(
\V{r}^{\prime }-\V{r})^{2}}{2\tau }\right]f_{L}(\V{r}^{\prime}) f_{R}(\V{r}) \,,
\end{eqnarray}
where $ \varphi_{LR}$ is the relative phase of the macroscopic wavefunctions defined in Eq. \eqref{eqn:varphid_Def}. With the change of variables $\bar{\V{r}}=\V{r}-\V{r}^{\prime }$, $\V{R}=(\V{r}+\V{r}^{\prime })/2$, we can rewrite Eq.~(\ref{Hom:Propposa}) as
\begin{eqnarray}
\label{bvorSaddle}
F_{L\rightarrow R}(t) = \Gamma\mathrm{Re}\,{\rm e}^{{\rm i} (\delta \mu t-\varphi_{LR})}
\int_{0}^{t}{\rm d}\tau \left( \frac{ {\rm i} \,m}{2\pi \hbar \tau }\right) ^{\frac{3}{2}}
{\rm e}^{{\rm i} (\omega _{\V{q}}-\Omega )\tau }  \int {\rm d}\bar{\V{r}}\int {\rm d}\V{r}  \exp \left[ \frac{{\rm i} }{\hbar }\frac{m\,}{2\tau }
\left( \bar{\V{r}}-\frac{\hbar \V{q}}{m}\tau
\right) ^{2}\right]  f_L(\V{R}-\frac{\bar{\V{r}}}{2})f_R(\V{R}+\frac{\bar{\V{r}}}{2}) \, .
\end{eqnarray}
\end{widetext}
The exponential in the integral over $\bar{\V{r}}$
oscillates very fast with respect to the wave functions $f_j(\V{r})$. Therefore 
the main contribution to the integral over $\bar{\V{r}}$
comes from $\bar{\V{r}}_0=\frac{\hbar \V{q}}{m}\tau $, where the
term in the exponential vanishes. By means of the saddle-point approximation
we take the wave functions at the point $\bar{\V{r}}_0 $  out of
the integral.
Performing the $\bar{\V{r}}$ integration using the Fresnel integral
\cite{Homodyne}
\begin{equation}
\int_{-\infty }^{\infty }{\rm d}t\,{\rm e}^{{\rm i} \gamma t^{2}}=\sqrt{\frac{\pi }{|\gamma
|}}{\rm e}^{{\rm i} {\rm sign}(\gamma )\pi /4} \, ,
\end{equation}
we obtain
\begin{eqnarray}\label{eqn:FCInt_pos}
F_{L\rightarrow R}(t) &\approx &\Gamma\mathrm{Re}\,{\rm e}^{{\rm i} (\delta \mu
t-\varphi_{LR})
}\int_{0}^{t}{\rm d}\tau
{\rm e}^{{\rm i} (\omega _{\V{q}}-\Omega )\tau } \nonumber \\ && \times
\int {\rm d}
\V{r}f_L\left(\V{r}+\V{d}-v_\V{q} \tau \right )f_R(\V{r}) \, , 
\end{eqnarray}
where $v_\V{q}$ is the recoil velocity, Eq. \eqref{recoil:v}.
The calculation that leads to Eq.~(\ref{eqn:FCInt_pos}) is exact in the limit
of homogeneous atomic systems.
In such case the momentum distribution of the condensate fraction is a
Dirac-delta function at zero momentum and all outcoupled atoms
have exactly the same momentum $\hbar \V{q}$. 
The condensates we consider are confined by an external potential and have a
finite extension in space which leads to a certain width $\delta p$ in their
momentum distribution, and thus to a spread in the momentum of the outcoupled atoms around the mean value $\hbar \V{q}$.  By taking the value of the atomic wave functions at the point of the stationary
phase of the exponential in Eq.~(\ref{bvorSaddle}) one neglects this momentum width:
The saddle-point approximation is thus applicable if $\hbar |\V{q}| \gg \delta p$.
Equation~(\ref{eqn:FCInt_pos}) agrees with the corresponding expression in
Eq.~(\ref{eqn:FCInt_posText}a). 
For completeness we also give the contributions of the macroscopic wavefunctions
to the background contribution, Eq.~(\ref{F:B}),
making the same approximations as in Eqs.~(\ref{eqn:FCInt_pos}): 
\begin{eqnarray}\label{eqn:FCB_pos}
 F_L(t) &\approx &\Gamma\mathrm{Re} \int_{0}^{t}{\rm d}\tau
{\rm e}^{{\rm i} (\omega _{\V{q}}-\Omega )\tau }\int {\rm d}
\V{r} f_L(\V{r})  f_L(\V{r}-v_\V{q}\tau ) \, ,\nonumber \\
F_R(t) &\approx &\Gamma\mathrm{Re}\, \int_{0}^{t}{\rm d}\tau {\rm e}^{{\rm i} (\omega
_{\V{q}}-\Omega +\delta \mu)\tau }\int \ {\rm d}\V{r}\,f_R(\V{r})f_R(
\V{r}-v_\V{q}\tau ) \, .\nonumber \\
\end{eqnarray}

\section{} \label{app:Ketterle}
We derive here approximate expressions of the Raman scattering
rate for the experimental setup of~\cite{Saba_DetPhase}, which lead to
Eq.~(\ref{eqn:Ftpostot}).
Using Eq.~(\ref{eqn:ThomasFermi}) for the condensate wavefunctions in Eq.~(\ref{eqn:FCB_pos}) we find 
\begin{eqnarray} \label{eqn:Fposcalc}
  F_{L}(t) 
&\approx&  \frac{2\pi \Gamma \mu(0) r_x \bar{R}^3}{g_1 v_\V{q}} \int_{0}^{z_<(0)}  {\rm d} z
\cos \left [ (\omega _{\V{q}}-\Omega)\frac{z r_x}{v_\V{q}} \right ]
G(z) \, ,\nonumber \\
\end{eqnarray}
with
\begin{eqnarray}
z_{<}(x) &=& {\rm  min}(2,\frac{v_\V{q}t-x}{r_x})\, .
\end{eqnarray}
The length $\bar{R}=(r_x r_y r_z)^{1/3}$ determines the typical size of the
condensates
and can be written as~\cite{Pethick_Book}
\begin{equation} \label{eqn:Def_Rbar}
 \bar{R}=15^{1/5}\left (\frac{Na_s}{\bar{a}} \right )^{1/5} \bar{a} \, .
\end{equation}
\begin{figure}[htp] \centering
\includegraphics[width=.44\textwidth]{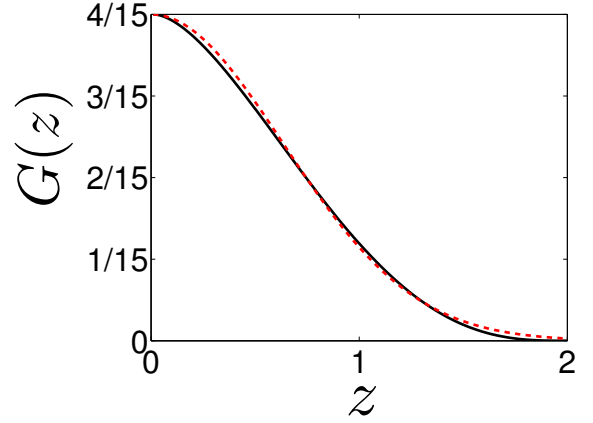} 
\caption{\label{fig:GaIntVgl}(color online) Comparison of the integral $G(z)$ as given in
Eq.~(\ref{eqn:GaInt}) (black solid line) with the Gaussian fit
Eq.~(\ref{eqn:Gz_gauss}) (red dashed line). }
\end{figure}
The function $G(z)$ in Eq.~(\ref{eqn:Fposcalc}) reads
\begin{eqnarray}\label{eqn:GaInt}
 G(z) &=& 2 \int_{z/2}^1 {\rm d}x \int_{0}^{\sqrt{1-x^2}} r {\rm d}r  \left (1-x^2-r^2
\right )^{1/2} \nonumber \\ && \times \left  (1-(x-z)^2-r^2\right )^{1/2} \, .
\end{eqnarray}
A numerical evaluation of $G(z)$ is shown in Fig.~(\ref{fig:GaIntVgl}), and is here compared
with a Gaussian of the form
\begin{equation} \label{eqn:Gz_gauss}
 G(z) \approx \frac{4}{15} e^{- 1.25 z^2} \, ,
\end{equation}
showing that this function provides a good approximation of Eq. \eqref{eqn:GaInt}.
Using Eq.~(\ref{eqn:Gz_gauss}) in Eq.~(\ref{eqn:Fposcalc})  one gets
\begin{eqnarray}\label{eqn:FluxesPosSmallTimes}
 F_L(t) &\approx &  \Gamma  N_C t_0 \int_0^{z_<(0)} {\rm d}z \cos \left [
(\omega_\V{q}-\Omega) z t_0 \right ] e^{- 1.25 z^2} \, , \nonumber \\
\end{eqnarray}
with $t_0=r_x/v_\V{q}$. For times $t>t_c$, with $t_c\sim (d-2r_x)/v_\V{q}$, such that $z_<(0)=z_<(d)=2$ we find from
Eq.~(\ref{eqn:FluxesPosSmallTimes}) 
\begin{eqnarray}\label{eqn:Err}
  F_{L}(t) &\approx & 
\Gamma t_0 N_C \int_{0}^2  {\rm d} z
\cos \left [ (\omega _{\V{q}}-\Omega)\frac{z r_x}{v_\V{q}} \right ]
e^{-1.25 z^2}  \nonumber \\
&\approx & \pi \Gamma N_C \sqrt{\frac{ t_0^2 }{5 \pi} } 
e^{-\frac{1}{5}[(\omega_{\V{q}}-\Omega)t_0]^2} \, .
\end{eqnarray}
When performing the integral in Eq.~(\ref{eqn:Err}) we neglected the imaginary
part in the Error-functions which one gets from the exact integration. This is a good
approximation
for  $|(\omega_{\V{q}}-\Omega)t_0|\le 5$, as can be checked by numerical
evaluation.  
For values $|(\omega_{\V{q}}-\Omega)t_0|> 5$  the exponential in
Eq.~(\ref{eqn:Err}) is negligible compared to the  resonant case
$\omega_{\V{q}}=\Omega$: The oscillating tails of the Error-functions given by their
imaginary parts will only play a role for parameters where the outcoupling
efficiency vanishes and which are thus not relevant to our treatment.
Using the same argumentation for the other contributions to the photon flux we
find
\begin{subequations} \label{eqn:FtC_position_final}
\begin{eqnarray}
 F_L(t) &\approx & \pi \Gamma N_C K(\omega_{\V{q}}-\Omega) \, , \\
 F_R(t) &\approx & \pi \Gamma N_C K(\omega_{\V{q}}-\Omega+\delta \mu) \, , \\
 F_{L\rightarrow R}(t) &\approx & 2 \pi \Gamma N_C K(\omega_{\V{q}}-\Omega) 
F_{\Theta}(q,t) \nonumber \\ & & \times \cos \left [\delta \mu
t-\varphi_{LR}+(\omega_{\V{q}}-\Omega )\frac{d}{v_{q}} \right ] \, ,\\
 F_{R \rightarrow L}(t) &\approx & 2 \pi \Gamma N_C
K(\omega_{\V{q}}-\Omega+\delta \mu) F_{\Theta}(-q,t) \nonumber \\ & & \times
\cos \left [\delta \mu t-\varphi_{LR}-(\omega_{\V{q}}-\Omega +\delta
\mu)\frac{d}{v_{q}} \right ] 
 \, ,\nonumber \\
\end{eqnarray}
\end{subequations}
with
\begin{eqnarray}
 K(x) &=& \sqrt{\frac{ t_0^2 }{5 \pi} }  e^{-\frac{t_0^2}{5} x^2} \, , \\ 
F_{\Theta}(q,t) &=& \Theta(q) \Theta \left(t-\frac{d-2r_x}{v_q}\right)       \,
.
\end{eqnarray}
From Eqs.~(\ref{eqn:FtC_position_final}) one then obtains
Eq.~(\ref{eqn:Ftpostot})
for $q>0$.
The phase difference between  $F_{L\rightarrow R}(t)$ and $F_{R \rightarrow L}(t)$
as measured in \cite{Shin_OptWeakLinkBEC} is obtained from Eqs.~(\ref{eqn:FtC_position_final}c,d)
and reads
\begin{equation} \label{eqn:PhaseOffset}
\Theta =\frac{d}{v_{q}}(2\Omega-2\omega _{\V{q}} - \delta \mu )
\, .
\end{equation}%

\begin{figure}[htp] \centering
\includegraphics[width=.44\textwidth]{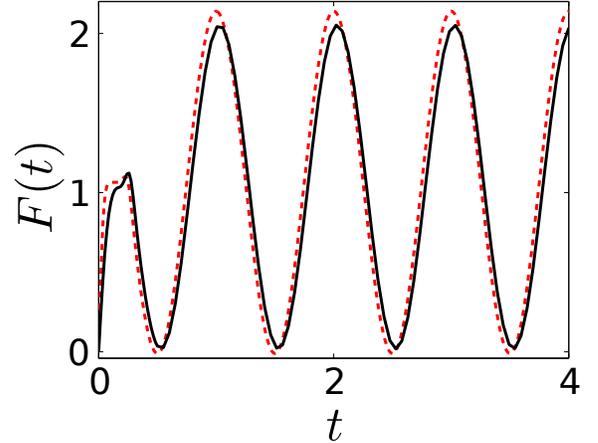} 
\caption{\label{fig:FluxesMomentumAlpha0} (color online)  Total photon Flux
$F(t)$ in units of background contribution $F_B(t)$ as a function of time in units
of $\frac{\delta \mu}{2\pi}$. The solid black line is computed by numerically
integrating Eq.~(\ref{eqn:FluxesMomentumKetterle}) and
 compared to the approximate result  Eq.~(\ref{eqn:Ftpostot}) (dashed red line).
  Parameters are as in Fig.~(\ref{fig:FluxesMomentum}) with $\alpha=0$.  }
\end{figure}

Figure~(\ref{fig:FluxesMomentumAlpha0}) displays the total photon flux computed 
from Eq.~(\ref{eqn:Ftpostot}) (dashed line) and the one calculated from
Eq.~(\ref{eqn:FluxesMomentumKetterle}) (solid line) for the experimental
parameters of \cite{Saba_DetPhase}. Quantitative agreement between the two
solutions is found, showing that neglecting the initial momentum distribution is
well justified for the considered parameters.

\end{appendix}


\begin{thebibliography}{99}

\bibitem{Stringari_RMP}
F. Dalfovo, S. Giorgini, L. P. Pitaevskii, and S. Stringari, Rev. Mod. Phys.
{\bf 71}, 463 (1999) 

\bibitem{Bloch-Review}
I. Bloch, J. Dalibard, and W. Zwerger, Rev. Mod. Phys. \textbf{80}, 885 (2008).

\bibitem{Demler_Lectures}
A. Imambekov, V. Gritsev, and E. Demler, "Fundamental noise in matter
interferometers", in "Ultracold Fermi gases", Proceedings of the International
School of Physics “Enrico Fermi”, 2006 (IOS Press, Amsterdam, The Netherlands,
2007). 

\bibitem{Bach}
R. Bach and K. Rza{\.z}ewski, Phys. Rev. Lett. {\bf 92}, 200401 (2004) 

\bibitem{Zwerger}
S. P. Rath and W. Zwerger, Phys. Rev. A {\bf 82}, 053622 (2010).

\bibitem{Greiner_Science_2010}
W. S. Bakr, A. Peng, M. E. Tai, R. Ma, J. Simon, J. I. Gillen, S. F\"olling, L.
Pollet, and M. Greiner, Science {\bf 329}, 547 (2010).

\bibitem{Bloch_Nature_2010}
J. F. Sherson, C. Weitenberg, M. Endres, M. Cheneau, I. Bloch, and S. Kuhr,
Nature {\bf 467}, 68 (2010); 
C. Weitenberg, M. Endres, J. F. Sherson, M. Cheneau, P. Schau\ss, T. Fukuhara,
I. Bloch, and S. Kuhr, Nature {\bf 471}, 319 (2011).

\bibitem{Greiner_PRL2011}
R. Ma, M. E. Tai, P. M. Preiss, W. S. Bakr, J. Simon, and M. Greiner, Phys. Rev.
Lett. {\bf 107}, 095301 (2011) 

\bibitem{Bloch_Science_2011}
M. Endres, M. Cheneau, T. Fukuhara, C. Weitenberg, P. Schau\ss, C. Gross, L.
Mazza, M. C. Ba{\~n}uls, L. Pollet, I. Bloch, and S. Kuhr, Science {\bf 334},
200 (2011). 

\bibitem{Reichel}
Y. Colombe, T. Steinmetz, G. Dubois, F. Linke, D. Hunger, and J. Reichel, Nature
{\bf 450}, 272 (2007).

\bibitem{Esslinger}
K. Baumann, C. Guerlin, F. Brennecke and T. Esslinger, Nature {\bf 464}, 1301
(2010);
K. Baumann, R. Mottl, F. Brennecke, and T. Esslinger, Phys. Rev. Lett. {\bf
107}, 140402 (2011).

\bibitem{Courteille}
S. Bux, C. Gnahm, R. A. W. Maier, C. Zimmermann, and Ph. W. Courteille, Phys.
Rev. Lett. {\bf 106}, 203601 (2011).

\bibitem{Stamper-Kurn}
K. W. Murch, K. L. Moore, S. Gupta, and D. M. Stamper-Kurn, Nature Phys. {\bf
4},  561 (2008); 
N. Brahms, T.P. Purdy, D.W.C. Brooks, T. Botter, and D.M. Stamper-Kurn, 
preprint arXiV:1012.1285 (2010), to appear in Nature Phys. (2011).

\bibitem{Meystre}
W. Chen, D. Meiser, and P. Meystre, Phys. Rev. A {\bf 75}, 023812 (2007).

\bibitem{RitschNature07}
I. B. Mekhov. C. Maschler and H. Ritsch , Nature Phys. {\bf 3}, 319 (2007).

\bibitem{LarsonPRL}
J. Larson, B. Damski, G. Morigi, and M. Lewenstein, Phys. Rev. Lett. {\bf 100},
050401 (2008).

\bibitem{Mekhov_QND}
I. B. Mekhov and H. Ritsch, Phys. Rev. Lett. {\bf 102}, 020403 (2009).

\bibitem{Hemmerich}
M. Weidem\"uller, A. Hemmerich, A. G\"orlitz, T. Esslinger, and T.W. H\"ansch,
Phys. Rev. Lett. {\bf 75}, 4583 (1995);
M. Weidem\"uller, A. G{\"o}rlitz, T.W. H{\"a}nsch, and A. Hemmerich, Phys. Rev.
A {\bf 58}, 4647 (1998).

\bibitem{Phillips}
G. Birkl, M. Gatzke, I. H. Deutsch, S. L. Rolston, and W. D. Phillips, Phys.
Rev. Lett. {\bf 75}, 2823 (1995).

\bibitem{Zimmermann_Bragg}
S. Slama, C. von Cube, A. Ludewig, M. Kohler, C. Zimmermann, and
Ph. W. Courteille, Phys. Rev. A {\bf 72}, 031402(R) (2005);
S. Slama, C. von Cube, B. Deh, A. Ludewig, C. Zimmermann, and
Ph.W. Courteille, Phys. Rev. Lett. {\bf 94}, 193901 (2005);
S. Slama, C. von Cube, M. Kohler, C. Zimmermann, and Ph. W.
Courteille, Phys. Rev. A {\bf 73}, 023424 (2006).

\bibitem{LightScattMottBloch}
C.~Weitenberg, P.~Schau\ss{}, T.~Fukuhara, M.~Cheneau, M.~Endres, I.~Bloch and
S.~Kuhr, Phys. Rev. Lett. {\bf 106} 215301 (2011)

\bibitem{Davidson_RMP}
R. Ozeri, N. Katz, J. Steinhauer, and N. Davidson, Rev. Mod. Phys. {\bf 77}, 187
(2005) 

\bibitem{Rist2010}
S. Rist, C. Menotti, and G. Morigi, Phys. Rev. A {\bf 81}, 013404 (2010).

\bibitem{Douglas11}
J. S. Douglas and K. Burnett, Phys. Rev. A {\bf 84}, 033637 (2011).

\bibitem{Lewenstein93}
M. Lewenstein and L. You, Phys. Rev. Lett. {\bf  71}, 1339 (1993). 

\bibitem{Javanainen}
J. Javanainen and J. Ruostekoski, Phys. Rev. Lett. {\bf 91}, 150404 (2003);
J. Ruostekoski, J. Javanainen, and G.V. Dunne, Phys. Rev. A {\bf 77}, 013603
(2008).

\bibitem{Ruotekoski09}
J. Ruostekoski, C.J. Foot, and A.B Deb, Phys. Rev. Lett. {\bf  103},
170404(2009).

\bibitem{Homodyne}
W.~P.~Schleich, {\it Quantum Optics in Phase Space}, Wiley ed. (New York, 2001).

\bibitem{Stringari_Book}
L.Pitaevski and S. Stringari, {\it Bose-Einstein Condensation} (Oxford Science
Publications, Oxford, 2003)

\bibitem{Saba_DetPhase}
M.~ Saba, T.~A.~Pasquini, C.~Sanner, Y.~Shin, W.~Ketterle and D.~E.~Pritchard, 
Science {\bf 307}, 1945 (2005)

\bibitem{Shin_OptWeakLinkBEC}
Y.~Shin, G.-B. Jo, M.~ Saba, T.~A.~Pasquini, W.~Ketterle and D.~E.~Pritchard. 
Phys.~Rev.~Lett. {\bf 95} 170402 (2005)

\bibitem{scully_QuantumEraser}
M.~O.~Scully and K.~Dr\"uhl, Phys.~Rev.~A {\bf 25}, 2208 (1982)


\bibitem{Maciej}
M. Lewenstein, L. You, J. Cooper, and K. Burnett, Phys. Rev. A {\bf 50}, 2207
(1994).

\bibitem{QTweezers} S. Zippilli, B. Mohring, E. Lutz, G. Morigi, and W. P.
Schleich, Phys. Rev. A {\bf 83}, 051602(R) (2011).
 
\bibitem{Larson_NJP}
J. Larson, S. Fernandez-Vidal, G. Morigi, and M. Lewenstein, New J. Phys. {\bf
10}, 045002 (2008).

\bibitem{comrel} This commutation relation is well defined
provided that ${\bf r},{\bf r'}\neq 0$. The mathematical subtlety related to
this specific point is irrelevant to our study, being practically zero the
probability that an atom in the electronic state $\ket{1}$ can be found around
this point in space.

\bibitem{LuxatOutcoupling}
D.~L.~Luxat and A.~Griffin, Phys. Rev. A {\bf 65}, 043618 (2002)

\bibitem{Choi_Outcoupling}
S.~Choi and Y.~Japha  and K.~Burnett, Phys. Rev. A {\bf 61}, 063606 (2000)

\bibitem{intout}  We note that the interaction between the
trapped and outcoupled atoms can be made small by means of a Feschbach
resonance \cite{Bloch-Review}.

\bibitem{menotti2008} C. Menotti and N. Trivedi, Phys. Rev. B 77, 235120 (2008).


\bibitem{corsright}
The corresponding expressions for the right cloud are obtained by
replacing $\Omega\rightarrow \Omega- \delta \mu$ where $\delta
\mu=\frac{\mu_L-\mu_R}{\hbar}$.


\bibitem{Cohen} C. Cohen-Tannoudji, J. Dupont-Roc, and G.
Grynberg, {\it Atom-Photon Interactions} (Wiley, New York, 2004).

\bibitem{Pitaevski99}
L. Pitaevskii and S. Stringari, Phys. Rev. Lett. {\bf 83}, 4237 (1999).

\bibitem{Janne}
J.~Roustekoski and D.~F.~Walls, Phys. Rev. A {\bf 56}, 2996 (1997).

\bibitem{Hadzibabic}
Z. Hadzibabic, S. Stock, B. Battelier, V. Bretin, and J. Dalibard, Phys. Rev. Lett. {\bf 93}, 180403 (2004).

\bibitem{Schmiedmayer}
S. Hofferberth, I. Lesanovsky, T. Schumm, A. Imambekov, V. Gritsev, E. Demler, and J. Schmiedmayer, Nature Physics {\bf 4}, 489 (2008);
A. Imambekov, I. E. Mazets, D. S. Petrov, V. Gritsev, S. Manz, S. Hofferberth, T. Schumm, E. Demler, and J. Schmiedmayer, Phys. Rev. A {\bf 80}, 033604 (2009).

\bibitem{Iazzi}
M. Iazzi and K. Yuasa, Phys. Rev. A {\bf 83}, 033611 (2011).

\bibitem{Mandel}
G. Magyar and L. Mandel, Nature (London) {\bf 198}, 255 (1963);
R. L. Pfleegor and L. Mandel, Phys. Rev. {\bf 159}, 1084 (1967).

\bibitem{Paul}
H. Paul, Rev. Mod. Phys. {\bf 58}, 209 (1986).

\bibitem{Englert_WhichWay}
B. G. Englert,  Phys.~Rev.~Lett. {\bf 77}, 2154 (1996)

\bibitem{Monroe}
L.-M. Duan and C. Monroe, Rev. Mod. Phys. {\bf 82}, 1209 (2010).


\bibitem{Priscilla} P. Ca\~nizares, T. G\"orler, J.P. Paz, G. Morigi, and W.P.
Schleich, Laser Physics {\bf 17}, 903 (2007).

\bibitem{CondDepletion}
We note that the condensate depletion, following from the outcoupling, induces a variation of $N_C$ as a function of time, which also leads to  a change in the chemical potential. This effect is not taken into account in our treatment, where we neglect the change of atom number in the condensates.

\bibitem{footcrittemp}
The critical temperature $T_c^i$ is
determined from the condition $N_C(T=T_c^i)=0$ and is evaluated numerically.  

\bibitem{Sachdev}
S.~Sachdev {\it Quantum Phase Transitions} (Cambridge University Press, 1999)

\bibitem{Spielman}
I.~B.~Spielman, W.~D.~Phillips and J.~V.~Porto, Phys. Rev. Lett. {\bf 98}, 080404 (2007).


\bibitem{CondFracPhillips}
I.~B.~Spielman, W.~D.~Phillips and J.~V.~Porto, Phys. Rev. Lett. {\bf 100}
120402 (2008).

\bibitem{Jaksch_BH}
D.~Jaksch, C.~Bruder, J.~I.~Cirac, C.~W.~Gardiner and P.~Zoller,
Phys. Rev. Lett. {\bf 81} 3108 (1998)

\bibitem{Kohn}
The use of the Gaussian ansatz in place of Wannier functions
should be handled with
care, for Wannier functions decay exponentially at infinity, see e.g. W. Kohn,
Phys. Rev. {\bf 115}, 809 (1959). Nevertheless, 
the precise form of the ansatz is not important for the purpose of this
estimate.

\bibitem{Wessel}
Here, we use the critical value determined by the mean-field, $r_c=3-2\sqrt{2}\approx 0.17$, see \cite{Sachdev}. We remark that a more precise determination of $r_c$ has been reported in S. Wessel, F. Alet, M. Troyer, and G. Batrouni, Phys. Rev. A {\bf 70}, 053615 (2004), where it was obtained by means of quantum Monte-Carlo simulations.

\bibitem{Duan}
L.-M. Duan, Phys. Rev. Lett. {\bf 96}, 103201 (2006).


\bibitem{CarusottoPRL2007}
T.~L.~Dao, A.~Georges, J.~Dalibard, C.~Salomon and J.~Carusotto, Phys. Rev.
Lett. {\bf 98} 240402 (2007).

\bibitem{SpatOrderBloch}
I.~Bloch, T.~W.~H\"ansch and T.~Esslinger, Nature, {\bf 403} 6766 (2000).

\bibitem{LongRangeEsslinger2007}
S.~Ritter, A.~\"Ottl, T.~Donner, T.~Bourdel, M.~K\"ohl and T.~Esslinger, Phys.
Rev. Lett, {\bf 98} 090402 (2007).

\bibitem{Gorshkov}
A. V. Gorshkov, L. Jiang, M. Greiner, P. Zoller, and M. D. Lukin, 
Phys. Rev. Lett. {\bf 100}, 093005 (2008).

\bibitem{Eschner}
J. Eschner, Ch. Raab, F. Schmidt-Kaler, and R. Blatt, Nature (London) {\bf 413}, 495 (2001).

\bibitem{Polzik}
K. Ekert, O. Romero-Isart, M. Rodriguez, M. Lewenstein, E. Polzik and A.
Sanpera, Nature Physics, {\bf 4} 50 (2008).

\bibitem{Chang}
D. Chang, V. Gritsev. G. Morigi, V. Vuletic, M.D. Lukin, and E.A. Demler, Nature physics {\bf 4}, 884 (2008).

\bibitem{Pethick_Book}
C.~J.~Pethick and H.~Smith {\it Bose Einstein Condensation in Dilute Gases}
(Cambridge University Press, 2008)


\end{thebibliography}
\end{document}